\documentclass[12pt]{article}
\usepackage{amsmath}
\usepackage{multirow}
\usepackage{amsfonts}
\usepackage{amssymb,graphics,psfrag}
\usepackage{array,epsfig,multirow,stmaryrd,graphicx}
\usepackage{comment}
\usepackage{slashed}
\usepackage{hyperref}
\usepackage{tensor}
\usepackage{booktabs}
\usepackage{colortbl}
\usepackage{hhline}
\usepackage{cite}

\def\hybrid{\topmargin -20pt    \oddsidemargin 0pt
        \headheight 0pt \headsep 0pt 
        \textwidth 6.25in      
        \textheight 9 in      
        \marginparwidth .875in
        \parskip 5pt plus 1pt
          \jot = 1.5ex
  }
\hybrid
\numberwithin{equation}{section}
\numberwithin{table}{section}

\newcommand{\beq}{\begin{equation}}
\newcommand{\eeq}{\end{equation}}
\newcommand{\be}{\begin{equation}}
\newcommand{\ee}{\end{equation}}
\newcommand{\bea}{\begin{eqnarray}}
\newcommand{\eea}{\end{eqnarray}}
\newcommand{\ben}{\begin{eqnarray*}}
\newcommand{\een}{\end{eqnarray*}}               
\newcommand{\ba}{\begin{aligned}}
\newcommand{\ea}{\end{aligned}}
\newcommand{\bt}{\begin{tabular}}
\newcommand{\et}{\end{tabular}}
\newcommand{\bc}{\begin{center}}
\newcommand{\ec}{\end{center}}

\newcommand{\cD}{\mathcal{D}}
\newcommand{\cL}{\mathcal{L}}

\newcommand{\cN}{\mathcal{N}}

\newcommand{\cH}{\mathcal{H}}

\newcommand{\cI}{\mathcal{I}}
\newcommand{\cJ}{\mathcal{J}}
\newcommand{\cR}{\mathcal{R}}

\newcommand{\cV}{\mathcal{V}}

\newcommand{\cM}{\mathcal M}

\newcommand{\I}{\text{Im}}
\newcommand{\R}{\text{Re}}

\newcommand{\bi}{{\bar \imath}}
\newcommand{\ib}{{\bar\imath }}
\newcommand{\jb}{{\bar\jmath }}
\newcommand{\bj}{{\bar\jmath}}

\DeclareMathOperator{\rk}{rank}
\DeclareMathOperator{\vol}{vol}
\DeclareMathOperator{\sign}{sign}

\newcommand{\nn}{\nonumber}

\newcommand{\cref}{{\bf [check ref]}}

\newcommand{\D}{\mathrm{D}}     

\psfrag{n1}{$\nu_1$}
\psfrag{n2}{$\nu_2$}
\psfrag{n1'}{$\nu_1'$}
\psfrag{n2'}{$\nu_2'$}
\psfrag{n9}{$\nu_9$}
\psfrag{n10'}{$\nu_{10}'$}
\psfrag{t1}{${\nu}_1$}
\psfrag{t2}{${\nu}_2$}
\psfrag{t9}{${\nu}_9$}
\psfrag{t1'}{${\nu}_1'$}
\psfrag{t2'}{${\nu}_2'$}
\psfrag{t10'}{${\nu}_{10}'$}

\newcommand{\BB}{{\boldsymbol{B}}}

\newcommand{\Bpsi}{{\boldsymbol{\psi}}}

\newcommand{\Blambda}{{\boldsymbol{\lambda}}}

\newcommand\Tstrut{\rule{0pt}{3.5ex}}     
\newcommand\Bstrut{\rule[-1.5ex]{0pt}{0pt}}
\newcommand\TTstrut{\rule{0pt}{2.3ex}}     
\newcommand\BBstrut{\rule[-0.9ex]{0pt}{0pt}}

\newcommand{\tiv}{\tilde{v}}
\newcommand{\dd}{d}

\def\blfootnote{\xdef\@thefnmark{}\@footnotetext}
\long\def\symbolfootnote[#1]#2{\begingroup%
\def\thefootnote{\fnsymbol{footnote}}\footnote[#1]{#2}\endgroup}

\begin{document}

\baselineskip=15pt

\begin{titlepage}
\begin{flushright}
\parbox[t]{1.8in}{\begin{flushright} MPP-2014-339 \end{flushright}}
\end{flushright}

\begin{center}

\vspace*{ 1.2cm}

{\large \bf  Partial Supergravity Breaking and\\[.2cm]
                   the Effective Action of Consistent Truncations}

\vskip 1.2cm

\begin{center}
 {Thomas W.~Grimm, Andreas Kapfer, and Severin L\"ust\ \footnote{grimm,\ kapfer,\ sluest@mpp.mpg.de}}
\end{center}
\vskip .2cm
\renewcommand{\thefootnote}{\arabic{footnote}}

{Max-Planck-Institut f\"ur Physik, \\
F\"ohringer Ring 6, 80805 Munich, Germany}

 \vspace*{1cm}

\end{center}

\vskip 0.2cm
 
 \begin{center} {\bf ABSTRACT } \end{center}
We study vacua of $\cN = 4$ half-maximal gauged supergravity in five dimensions
and determine crucial properties of the effective theory around the vacuum.
The main focus is on configurations with exactly two broken supersymmetries, 
since they frequently appear in consistent truncations of string theory and supergravity.
Evaluating one-loop corrections to the Chern-Simons terms we find necessary conditions 
to ensure that a consistent truncation also gives rise to a proper effective action of an 
underlying more fundamental theory. 
To obtain concrete examples, we determine the $\cN=4$ action of M-theory on six-dimensional 
$SU(2)$-structure manifolds with background fluxes. 
Calabi-Yau threefolds with vanishing Euler number are examples of $SU(2)$-structure manifolds 
that yield $\cN=2$ Minkowski vacua. We find that that one-loop corrections to the Chern-Simons terms vanish trivially 
and thus do not impose constraints on identifying effective theories. This 
result is traced back to the absence of isometries on these geometries. 
Examples with isometries arise from type IIB supergravity on squashed Sasaki-Einstein manifolds.
In this case the one-loop gauge Chern-Simons terms vanish due to non-trivial cancellations, while the one-loop 
gravitational Chern-Simons terms are non-zero.

\vskip 0.4cm

\hfill {September, 2014}
\end{titlepage}

\tableofcontents

\newpage



\section{Introduction}
Since the early days of supersymmetric field theory people have been studying spontaneous supersymmetry breaking.
It is often not only a necessity of phenomenological application but also offers insights into
properties of these theories in general.
While generically all supersymmetries may be broken,
vacua that preserve exactly half of them are of particular interest \cite{Cecotti:1984rk,Cecotti:1984wn,Cecotti:1985sf,Ferrara:1995gu,Antoniadis:1995vb,
Fre:1996js,Kiritsis:1997ca,Andrianopoli:2002rm,Louis:2009xd,Louis:2010ui,Hansen:2013dda,Hohm:2004rc,Grimm:2014soa}.
In \cite{Grimm:2014soa} it was found that the $\cN =4 \rightarrow \cN =2$ breaking of five-dimensional gauged supergravity can be triggered
by a St\"uckelberg-like mechanism, where tensors become massive by eating a vector.
In this paper we generalize this analysis and
use partial supersymmetry breaking as a tool to investigate consistent truncations of supergravity and string theory.

In principle for a general compactification of some higher dimensional theory on a compact manifold one has 
to include all massive and massless modes in the derivation of the effective action. 
In contrast, consistent truncations describe the dynamics only for a subset of all these modes.
By definition these modes are chosen such that solutions of the lower-dimensional equations of motion 
lift to solutions of the higher-dimensional equations of motion. 
It is this property that allows to use the truncated theories as tools for constructing higher-dimensional solutions.
However, recently consistent truncations were also used for phenomenology in non-Calabi-Yau compactifications.
Consequently, the effective action derived from a consistent truncation should better 
match the genuine effective action with the whole tower of massive modes integrated out.
Setups with partial supergravity breaking now allow us to 
derive necessary conditions for this agreement in theories
where we already know parts of the effective action,
like e.g.~Calabi-Yau compactifications. Here we investigate this issue in the context of one-loop corrections 
to the Chern-Simons terms. These are induced by massive charged modes at one-loop, but are 
nevertheless independent of the mass scale. Due to their topological nature and their relation to anomalies
they are very robust quantities that can contain non-trivial information about the massive spectrum.

In this paper we first study $\cN =2$ vacua of $\cN=4$ gauged supergravity in five dimensions using the embedding tensor formalism of
\cite{Schon:2006kz}, which encodes the gauging of global symmetries in a very convenient way.
After assigning vacuum expectation values (VEVs) to the scalars
we calculate the gravitino masses, i.e.~the number of broken supersymmetries, the cosmological constant, the bosonic spectrum including mass terms and charges,
as well as Chern-Simons terms. These quantities depend on the form of the embedding tensors contracted with the VEVs of the coset representatives
of the scalar manifold.
Given these objects one can fully analyze the theory around the vacuum. 
While such an analysis is possible for each considered vacuum, 
a classification of allowed vacua is beyond the scope of this paper.

As an application we then make contact with M-theory compactifications on $SU(2)$-structure manifolds.
First we study general consistent truncations of M-theory on $SU(2)$-structure manifolds to
$\cN=4$ gauged supergravity, before we restrict to the special case of
Calabi-Yau manifolds with vanishing Euler number, which have $SU(2)$-structure as well, as can be seen by the Poincar\'e-Hopf theorem.
These spaces constitute an $\cN=2$ Minkowski vacuum of the general $\cN =4$ gauged supergravity, including massive modes.
The same analysis has been carried out for the type IIA case in \cite{Danckaert:2011ju,KashaniPoor:2013en}.
Since the Chern-Simons terms in the genuine effective action of M-theory on a smooth Calabi-Yau threefold
are not corrected by integrating out massive modes \cite{Cadavid:1995bk,Papadopoulos:1995da,Ferrara:1996hh}, 
we demand that one-loop Chern-Simons terms should also be absent in the effective action 
of a consistent truncation.
For the analyzed example of the Enriques Calabi-Yau
it turns out that the massive modes are not charged under any massless vector,
and one-loop corrections therefore trivially cancel. 
This is one possible way to ensure that a consistent truncations on $SU(2)$ structure threefolds 
that are also Calabi-Yau can be compatible with the genuine effective action.  
However, already in the considered consistent truncation for the Enriques Calabi-Yau we miss at the 
massless level a vector multiplet and a hypermultiplet that are not captured by our particular $SU(2)$-structure 
ansatz. Nevertheless, we argue that one can consistently complete the 
Chern-Simons terms including an additional massless vector.

As a second example we consider
a particular consistent truncation of type IIB supergravity on a squashed Sasaki-Einstein
manifold with RR-flux. 
This is again described by five-dimensional $\cN =4$ gauged supergravity, and indeed there are $\cN=2$
vacua that are now AdS \cite{Cassani:2010uw,Liu:2010sa,Gauntlett:2010vu}.\footnote{See also \cite{Tsikas:1986rx,Lu:1999bw,Ceresole:1999zs,Cvetic:2000nc,Hoxha:2000jf,Buchel:2006gb,Gauntlett:2007ma,Skenderis:2010vz,Cassani:2010na,Bena:2010pr,
Bah:2010cu,Liu:2010pq,Liu:2011dw,Halmagyi:2011yd}
for related works on this subject.}
The most prominent example is certainly the five-sphere, although our results hold for any squashed Sasaki-Einstein manifold.
In the theory around the vacuum there are massive states that are charged under the gauged $U(1)$ R-symmetry.
Remarkably their one-loop corrections to the gauge Chern-Simons term cancel non-trivially, while the gravitational Chern-Simons term 
does receive corrections. While we are not able to give a precise interpretation of this fact, it is 
an intriguing observation that such cancellations take place. Let us also stress that in 
this AdS case the existence of an effective theory can be generally questioned, since 
the AdS radius might be linked to the size of the compactification space. 
It is not hard to see that the squashed Sasaki-Einstein reductions of type IIB are reminiscent of the general $SU(2)$ structure
reductions of M-theory considered in the first part of the paper. It was indeed argued that 
there is a relation between these two settings when using T-duality \cite{Gauntlett:2004zh,Gauntlett:2006ai,Gauntlett:2007sm,OColgain:2011ng,Sfetsos:2014tza},
if one includes warping in the $SU(2)$-structure ansatz, which is in general quite difficult and beyond the scope of this paper.

The paper is organized as follows: In \autoref{N=4Generalities} we review $\cN=4$ gauged supergravity in five dimensions using the embedding tensor formalism
and calculate the spectrum as well as the relevant parts of the Lagrangian around the vacuum. We proceed in \autoref{su2reduction}
with the general description of M-theory compactifications on $SU(2)$-structure manifolds. In \autoref{sec:ex}, after
stating some general remarks about the quantum effective action of consistent truncations, we analyze M-theory on the Enriques Calabi-Yau
and type IIB supergravity consistent truncation on a squashed Sasaki-Einstein
manifold.

\section{Gauged \texorpdfstring{$\cN =4$}{N=4} supergravity in five dimensions and its vacua} \label{N=4Generalities}

In this section we begin with some important facts about $\cN=4$ gauged supergravity theories in \autoref{N=4Gen}. 
In \autoref{sec:iso} we provide a tool to extract the propagating degrees of freedom out of the theory, since the standard formulation
in \cite{Schon:2006kz} uses vectors and dual tensors on equal footing.
Finally we study the vacua of this setup in \autoref{sec:action}.
We derive the mass terms and charges of the scalar and tensor fields and give expression for the
vector masses, field strengths and Chern-Simons terms. The results depend on the precise form of the embedding tensors
contracted with the scalar field VEVs.
Since we are in particular interested in preserving half of the supersymmetry in the vacuum, we also derive the mass terms of the gravitinos in
terms of the contracted embedding tensors.

\subsection{Generalities} \label{N=4Gen}

We start with a review of the general properties of $\cN =4$ gauged supergravity in five dimensions along the lines of
\cite{Dall'Agata:2001vb,Schon:2006kz}.\footnote{We stress that in our conventions
five-dimensional $\cN =4$ supergravity theories have 16 supercharges and thus are half-maximal supergravities.} First consider 
ungauged Maxwell-Einstein supergravity, which couples 
$n$ vector multiplets to a gravity multiplet. Note that as long as the theory is not gauged, one can replace the
vector multiplets by dual tensor multiplets. 
The gravity multiplet has the field content
\begin{align} \label{grav_multiplet}
 (g_{\mu\nu}, \psi^i_\mu , A^{ij}_\mu , A^0 , \chi^i , \sigma )\, ,
\end{align}
with the metric $g_{\mu\nu}$, four spin-3/2 gravitini $\psi^i_\mu$, six vectors $(A^{ij}_\mu , A^0)$, four spin-1/2 fermions $\chi^i$
and one real scalar $\sigma$. The indices of the fundamental representation of
the R-symmetry group $USp(4)$ are written as $i,j = 1,\dots ,4$. The symplectic form of $USp(4)$, denoted $\Omega$, has the following properties
\begin{align}\label{e:properties_omega}
 \Omega_{ij} = - \Omega_{ji} \, , \qquad \Omega_{ij} = \Omega^{ij} \, , \qquad \Omega_{ij}\Omega^{jk} = - \delta^k_i \, .
\end{align}
Raising and lowering of $USp(4)$ indices is done according to
\begin{align}\label{e:raising_lowering}
 V^i = \Omega^{ij} V_j \, , \qquad V_i = V^j \Omega_{ji} \, .
\end{align}
The double index $ij$ labels the \textbf{5} representation of $USp(4)$ with the following properties
\begin{align}
 A^{ij}_\mu = - A^{ji}_\mu \, , \qquad A^{ij}_\mu \, \Omega_{ij} = 0 \, , \qquad (A^{ij}_\mu)^* = A_{\mu \, ij} \, .
\end{align}
Since $USp(4)$ is the spin group of $SO(5)$, we will often use the local isomorphism
$SO(5) \cong USp(4)$ to switch between representations of both groups. The indices of
the fundamental representation of $SO(5)$ are denoted by $m,n = 1, \dots , 5$ and the Kronecker delta $\delta_{mn}$ is used to raise and lower them.
Moreover all massless fermions in this papers are supposed to be symplectic Majorana
spinors. For further conventions and useful identities consult \autoref{App-Conventions}.
Finally we will often use the definition
\begin{align}
 \Sigma := e^{\sigma / \sqrt 3}\, ,
\end{align}
where $\sigma$ is the real scalar of the gravity multiplet \eqref{grav_multiplet}.

To the gravity multiplet we can now couple $n$ vector multiplets, labeled by $a,b = 6, \dots , 5+n$. They are again raised
and lowered with the Kronecker delta $\delta_{ab}$. The multiplets read
\begin{align}\label{e:content_vector}
 (A_\mu^a , \lambda^{ia} , \phi^{ija})\, ,
\end{align}
where $A_\mu^a$ denote the vectors, $\lambda^{ia}$ spin-1/2 fermions and the $\phi^{ija}$ scalars in the \textbf{5} of $USp(4)$.

The set of all scalars in the theory span the manifold 
\begin{align} \label{coset_def}
 \cM = \cM_{5,n} \times SO(1,1) \, , \qquad \quad \cM_{5,n} = \frac{SO(5,n)}{SO(5)\times SO(n)}\, ,
\end{align}
where we parametrize the coset $\cM_{5,n}$ by the scalar fields $\phi^{ija}$ in the vector multiplets, whereas
the $SO(1,1)$ part is captured by the scalar $\sigma$ in the gravity multiplet. Hence the global symmetry group of the theory is found to be
$SO(5,n)\times SO(1,1)$.
We now define $SO(5,n)$ indices $M,N= 1,\dots , 5+n$, which we can
raise and lower with the $SO(5,n)$ metric $(\eta_{MN}) = \textrm{diag}(-1,-1,-1,-1,-1,+1,\dots ,+1)$.
The coupling of the vector multiplets to the gravity multiplet is realized by noting that all vectors in the theory transform 
as a singlet $A^0$ and the fundamental
representation $A^M$ of $SO(5,n)$:
\begin{align}
 (A^0 , A^{ij} , A^n ) \rightarrow (A^0 , A^{M})
\end{align}
and have $SO(1,1)$ charges $-1$ and $1/2$ for $A^0$ and $A^M$, respectively.
In the representation of the vector fields the generators $t_{MN}$ of $SO(5,n)$ and $t_0$ of $SO(1,1)$ read
\footnote{All antisymmetrizations in this paper include a factor of $1/n!\,$.}
\begin{align}
 \tensor{t}{_M_N_\, _P^Q} = 2\delta_{[M}^Q \, \eta_{N]P} \, , \qquad \tensor{t}{_0_\, _M^N} = - \frac{1}{2} \,\delta_M^N \, ,
 \qquad \tensor{t}{_M_N_\, _0^0} = 0 \, , \qquad \tensor{t}{_0_\, _0^0} = 1 \, .
\end{align}

The most convenient way to describe the coset space $\cM_{5,n}$ is via the coset representatives
$\cV=(\tensor{\cV}{_M^m} , \tensor{\cV}{_M^a})$, here $m = 1, \dots , 5$ and $a = 6, \dots n+5$
are the indices of the fundamental representations of $SO(5)$ and $SO(n)$, respectively.
The definition is such that local $SO(5)\times SO(n)$ transformations act from the right, while global $SO(5,n)$ transformations on $\cV$ act from the left.
It is important to notice that
\begin{align}\label{e:SO(5,N)_index_raising}
 \tensor{\cV}{_M^a} = \eta_{MN}\,\cV^{Na} \, , \qquad \tensor{\cV}{_M^m} = - \eta_{MN}\, \cV^{Nm} \, 
\end{align}
and also that, since
$(\tensor{\cV}{_M^m} , \tensor{\cV}{_M^a})\in SO(5,n)$, we have
\begin{align} \label{eta_viaV}
 \eta_{MN} = - \tensor{\cV}{_M^m} \cV_{N m} + \tensor{\cV}{_M^a} \cV_{N a} \, .
\end{align}
Furthermore we define a non-constant positive definite metric on the coset
\begin{align} \label{M_viaV}
 M_{MN} = \tensor{\cV}{_M^m} \cV_{N m} + \tensor{\cV}{_M^a} \cV_{N a} \, 
\end{align}
with inverse given by $M^{MN}$.
Lastly we introduce
\begin{align}
 M_{MNPQR} = \varepsilon_{mnpqr}\tensor{\cV}{_M^m}\tensor{\cV}{_N^n}\tensor{\cV}{_P^p}\tensor{\cV}{_Q^q}\tensor{\cV}{_R^r}\, ,
\end{align}
where $\varepsilon_{mnpqr}$ is the (flat) five-dimensional Levi-Civita tensor.

We proceed with the gauging of global symmetries. The different gaugings are most conveniently described using the
embedding tensors $f_{MNP}$, $\xi_{MN}$ and $\xi_M$, which are totally antisymmetric. They determine the covariant derivative\footnote{Note that 
a gauge coupling constant $g$ can explicitly be included whenever an embedding tensor appears. However for simplicity we take $g=1$ in the following.}
\begin{align} \label{gen_cov_der}
 D_\mu = \nabla_\mu -  A_\mu^M \tensor{f}{_M^N^P} t_{NP} -  A_\mu^0 \, \xi^{NP} t_{NP} -  A^M_\mu \xi^N t_{MN} -  A_\mu^M \xi_M t_{0}\, .
\end{align}
Note that in the ungauged theory the embedding 
tensors are supposed to transform under the global symmetry group.
Fixing a value for the tensor components, the global symmetry group is then broken down to a subgroup.
In this paper we will mostly set $\xi_{M}=0$, since the calculations simplify considerably and several interesting cases are already covered.
However a non-vanishing $\xi_{M}$ might then be included straightforwardly.
Accordingly the covariant derivative \eqref{gen_cov_der} simplifies to
\begin{align}
 D_\mu = \nabla_\mu -  A_\mu^M \tensor{f}{_M^N^P} t_{NP} - A_\mu^0 \, \xi^{MN} t_{MN} \, .
\end{align}
The embedding tensors satisfy $f_{MNP}=f_{[MNP]}$, $\xi_{MN}=\xi_{[MN]}$ and are, 
in the case of $\xi_M = 0$, subject to the quadratic constraints
\begin{align}\label{e:quadr_constr}
 &f_{R[MN}\tensor{f}{_P_Q_]^R}=0\, , & \tensor{\xi}{_M^Q}f_{QNP} = 0 \, .
\end{align}
For vanishing $\xi_M$ the linear constraints are trivially satisfied.
There is an important issue with such nontrivial gaugings, which forces us to dualize
some of the vector fields $A_\mu^M$ into two-forms $B_{\mu\nu\, M}$. Therefore we consider 
an action where both $A_\mu^M$ and $B_{\mu\nu\, M}$ are present in order
to write down a general gauged supergravity with $\xi_M = 0$.\footnote{As long as $\xi_M$ vanishes, we do not have to introduce a tensorial counterpart
$B^0_{\mu\nu}$ for $A^0_\mu$.}
Using this approach, the tensor fields $B_{\mu\nu\, M}$ carry no on-shell degrees of freedom.
However, they can eat a dynamical vector with three degrees of 
freedom and become massive. This will be treated in \autoref{sec:iso}.

The bosonic Lagrangian of this $\cN=4$ gauged supergravity theory is given by \cite{Dall'Agata:2001vb,Schon:2006kz}
\begin{align} \label{bos_N=4action}
 e^{-1}\cL_{\textrm{bos}}=&
 -\frac{1}{2}R - \frac{1}{4}\Sigma^2 M_{MN}\, \cH^M_{\mu\nu} \cH^{\mu\nu\, N} - \frac{1}{4}\Sigma^{-4}F_{\mu\nu}^0 F^{\mu\nu\, 0}\nn \\
 &-\frac{3}{2}\Sigma^{-2}(\nabla_\mu \Sigma)^2 
  + \frac{1}{16}(D_\mu M_{MN})(D^\mu M^{MN})\nn \\
 &+\frac{1}{16\sqrt 2}\epsilon^{\mu\nu\rho\lambda\sigma} \xi^{MN}B_{\mu\nu\, M} \big( D_\rho B_{\lambda\sigma\, N} 
 +4 \eta_{NP}A^0_\rho \partial_\lambda A^P_\sigma 
 + 4 \eta_{NP}A^P_\rho \partial_\lambda A^0_\sigma \big) \nn \\
 & - \frac{1}{\sqrt 2}\epsilon^{\mu\nu\rho\lambda\sigma}  A_\mu^0 \, \Big( \partial_\nu A^M_\rho \partial_\lambda A_{\sigma\, M}
 + \frac{1}{4}\xi_{MN} A^M_\nu  A^N_\rho \partial_\lambda A^0_\sigma
 - f_{MNP}  A^M_\nu A^N_\rho \partial_\lambda A^P_\sigma \Big ) \nn \\
 & -\frac{1}{4} f_{MNP}\,f_{QRS} \,\Sigma^{-2}\,\Big( \frac{1}{12}M^{MQ}M^{NR}M^{PS} - \frac{1}{4}M^{MQ}\eta^{NR}\eta^{PS} 
 + \frac{1}{6}\eta^{MQ}\eta^{NR}\eta^{PS}\Big)\nn\\
 &- \frac{1}{16} \xi_{MN}\,\xi_{PQ} \,\Sigma^4 \, \Big( M^{MP}M^{NQ}- \eta^{MP}\eta^{NQ} \Big) - \frac{1}{6\sqrt 2} f_{MNP}\, \xi_{QR}\,\Sigma\, M^{MNPQR} \, ,
\end{align}
where $R$ denotes the Ricci scalar, and we define
\begin{align} \label{def-cH}
 \cH^M_{\mu\nu} := 2\, \partial_{[\mu} A^M_{\nu]} -  \tensor{\xi}{_N^M}A_\mu^0 A_\nu^N - \tensor{f}{_P_N^M}A_\mu^P A_\nu^N +
 \frac{1}{2}\, \xi^{MN} B_{\mu\nu\, N}\, ,
\end{align}
as well as
\begin{align}
 F^0_{\mu\nu} := \partial_\mu A^0_\nu - \partial_\nu A^0_\mu \, .
\end{align}
The vectors and dual tensors are subject to
vector gauge transformations with scalar parameters $(\Lambda^0 , \Lambda^M)$ as well as
standard two-form gauge transformations with one-form parameters $\Xi_{\mu\, M}$. These
transformations will be of importance later, since they allow us to remove some of the vectors by
gauge transformations. For our choice of gaugings the variation of the vectors reads
\begin{align} \label{general_gaugetransform}
 \delta A^0_\mu = \nabla_\mu \Lambda^0 \, , \qquad \delta A^M_\mu = D_\mu \Lambda^M -\frac{1}{2} \xi^{MN} \Xi_{\mu\, N}\, .
\end{align}

We continue with the fermionic Lagrangian. 
To simplify our notation we introduce contractions of the embedding tensor
with the coset representatives
\begin{align}\label{e:dressed_gaugings}
 &\xi^{mn} := \tensor{\cV}{_M^m} \tensor{\cV}{_N^n}\, \xi^{MN} \, , \qquad \xi^{ab} := \tensor{\cV}{_M^a} \tensor{\cV}{_N^b} \,\xi^{MN}  \, , \qquad
 \xi^{am} := \tensor{\cV}{_M^a} \tensor{\cV}{_N^m} \,\xi^{MN}  \, , \nn \\
 &f^{mnp} := \tensor{\cV}{_M^m}  \tensor{\cV}{_N^n} \,\tensor{\cV}{_P^p} f^{MNP} \, , \qquad
 f^{mna} := \tensor{\cV}{_M^m}  \tensor{\cV}{_N^n} \,\tensor{\cV}{_P^a} f^{MNP} \, , \qquad \dots \, .
\end{align}
Note that these objects are field-dependent and acquire a VEV in the vacuum. It is important to realize that
the positions of the $SO(5,n)$-indices $M,N$ in \eqref{e:dressed_gaugings} are essential because of \eqref{e:SO(5,N)_index_raising}.
Using this notation we define
\begin{align}\label{e:shift_matrices}
\textbf{M}_\psi^{ij} := \textbf{M}_\psi^{mn} \,\tensor{\Gamma}{_m_n^i^j}\,  
\end{align}
with
\begin{align}
 & \textbf{M}_\psi^{mn} := - \frac{1}{4\sqrt 2}\Sigma^2 \,\xi^{mn} + \frac{1}{24}\epsilon^{mnpqr}\,f_{pqr} \, , &\Gamma_{mn} := \Gamma_{[m}\Gamma_{n]} \, ,
\end{align}
where $\Gamma_m$ are the $SO(5)$ gamma matrices.
We are now in the position to write down the relevant fermionic terms in the Lagrangian.
For the purpose of this work 
we will find it sufficient to only recall the kinetic terms and the mass terms of the gravitini.
The remaining quadratic terms of the fermions can be found in \cite{Dall'Agata:2001vb,Schon:2006kz}.
The relevant part of the Lagrangian reads
\begin{align} \label{fermionic-quadratic}
 e^{-1}\cL_{\textrm{ferm}}=&
 -\frac{1}{2} \bar \psi^i_\mu \,\gamma^{\mu\nu\rho}\,\cD_\nu\, \psi_{\rho\, i} 
 +\frac{1 }{2}i\, \textbf{M}_{\psi\, ij}\, \bar \psi^i_\mu \,\gamma^{\mu\nu} \,\psi^j_\nu \, .
\end{align}
The precise form of the covariant derivative is of no importance in this paper, since we will derive only the charges of the bosons in the vacuum
and infer the remaining ones by supersymmetry.
This concludes our discussion of the general properties $\cN =4$ gauged supergravity in five dimensions.

 \subsection{Isolation of the propagating degrees of freedom}\label{sec:iso}
The formulation of $\cN=4$ gauged supergravity in terms of embedding tensors, presented in \cite{Schon:2006kz},
is a very powerful way to implement general gaugings
of global symmetries.
However, in order to study supersymmetry breaking vacua and the resulting 
effective field theories we need to eliminate non-propagating degrees of freedom used in the democratic formulation of 
 \cite{Schon:2006kz}. In particular, the $\cN=4$ gauged supergravities  
 are formulated in terms of vectors and dual tensors.
 We eliminate redundant degrees of freedom in vectors by tensor gauge transformations,
rendering the corresponding tensors
the (massive) propagating degrees of freedom. All remaining tensors that are not involved 
in this gauging procedure turn out to decouple in the action and can be consistently set to zero.
In these cases the corresponding vectors constitute the appropriate formulation.
In the following we carry out the necessary redefinition of vectors and tensors explicitly.

The isolation of the appropriate propagating degrees of freedom in $\cN=4$ gauged 
supergravity depends on the form of the embedding tensor
$\xi^{MN}$.\footnote{We again stress that we set $\xi_M =0$ unless stated differently.}
This can easily be seen as follows. Consider the gauge transformations of the vectors $A^M$ \eqref{general_gaugetransform}
as well as the variation of the action with respect to the tensors $B_{\mu\nu\,M}$
\begin{align}
 &\delta A^M_\mu = D_\mu \Lambda^M -\frac{1}{2} \xi^{MN} \Xi_{\mu\, N} \, , 
 & \frac{\delta S}{\delta B_{\mu\nu\,M}} \sim \xi^{MN} (\dots)_N \, .
\end{align}
Note that one can always find orthogonal transformations such that
\begin{align}\label{e:full_rank_rot}
 &(\xi^{MN}) \mapsto \left( \begin{array}{c|c}
 \xi^{\hat M \hat N} &  \textbf{0}^{\hat M \tilde N} \Bstrut\\  \hline 
  \textbf{0}^{\tilde M \hat N} & \textbf{0}^{\tilde M \tilde N} \Tstrut
                           \end{array}\right) \, , \\
 & \hat M , \hat N =1,\dots , \rk (\xi^{MN}) \, , \qquad \tilde M , \tilde N =\rk (\xi^{MN}) + 1,\dots , 5+n \, , \nn
\end{align}
with $(\xi^{\hat M \hat N})$ a full-rank matrix. It is now easy to see that after appropriate partial gauge fixing
one can invert $(\xi^{\hat M \hat N})$ to obtain
\begin{align}
 \delta A^{\hat M}_\mu = - A^{\hat M}_\mu \, .
\end{align}
The $A^{\hat M}_\mu$ are therefore pure gauge and can be removed from the action. The corresponding tensors $B_{\mu\nu\,\hat M}$
constitute the appropriate formulation. In contrast, we find for the remaining vectors and tensors
\begin{align}
 &\delta A^{\tilde M}_\mu = D_\mu \Lambda^{\tilde M} \, , 
 & \frac{\delta S}{\delta B_{\mu\nu\,\tilde M}} = 0 \, .
\end{align}
The Lagrangian is therefore independent of the $B_{\mu\nu\,\tilde M}$, which is why we can remove them. We are left with
propagating vectors $A^{\tilde M}$ subject to standard vector gauge transformations. So we see that the propagating
degrees of freedom are captured by $A^{\tilde M}_\mu$, $B_{\mu\nu\,\hat M}$.
Moreover, for the pair $B^0_{\mu\nu}$, $A^0_\mu$ it turns out that the tensor $B^0_{\mu\nu}$ does not appear in the action
and $A^0_\mu$ constitutes the field carrying the propagating degrees of freedom.

Note that this procedure easily generalizes, if one allows for a non-vanishing $\xi_M$. In this case one just has to replace
$\xi^{MN}$ by $2 \cdot Z^{\cM\cN}$ in the previous calculations, where $\cM = (0,M)$ and
\begin{align}
 &Z^{MN} = \frac{1}{2} \xi^{MN} \, , & Z^{0M} = - Z^{M0} = \frac{1}{2} \xi^M \, .
\end{align}
One can then rotate $Z^{\cM\cN}$ into a full-rank part and zero-matrices as in \eqref{e:full_rank_rot}.
The fields $A^0_\mu$ and $B_{0\, \mu\nu}$ must then also be included in the procedure.
As already mentioned, we nevertheless set $\xi_M = 0$ in the following.

In this paper we are interested in deriving the half-supersymmetric
Lagrangian around a vacuum of the $\cN =4$ theory. In order to extract the propagating fields
we therefore slightly modify the approach we just described. This proves convenient for our purposes.
We start with the democratic formulation
of $\cN =4$ gauged supergravity reviewed in \autoref{N=4Gen} including redundancies. 
We than assume that we have found a vacuum in which  all scalars,
i.e.~$\langle \tensor{\cV}{_M^m} \rangle$, $\langle \tensor{\cV}{_M^a} \rangle$, $\langle \Sigma \rangle$, 
acquire a VEV. 
In analogy to \eqref{e:dressed_gaugings} we define
\begin{align} \label{rotate_AB}
&B_{\mu \nu}^m  := \tensor{\langle \cV\rangle}{_M^m} B_{\mu \nu}^M \, ,  
&& B_{\mu \nu}^a := \tensor{\langle\cV\rangle}{_M^a} B_{\mu \nu}^M \, , & \\
  &A_\mu^m := \tensor{\langle \cV\rangle}{_M^m} A_\mu^M \, ,  && A_\mu^a :=  \tensor{\langle\cV\rangle}{_M^a} A_\mu^M \, . & \nn
\end{align}
Similarly we can introduce the gauge parameters $(\Lambda^m,\Lambda^a)$ and
$(\Xi_{\mu}^m,\Xi_{\mu}^a )$ by setting 
\begin{align}
   &\Lambda^m:=  \tensor{\langle \cV\rangle}{_M^m} \Lambda^M  \,  , & & \Lambda^a  :=  \tensor{\langle  \cV\rangle}{_M^a}  \Lambda^M \, ,& \\
   &\Xi_{\mu}^m :=   \tensor{\langle \cV\rangle}{_M^m} \Xi_{\mu}^M  \, , &&  \Xi_{\mu}^a :=  \tensor{\langle\cV\rangle}{_M^a}  \Xi_{\mu}^M \, .&    \nn
\end{align}
In this rotated basis
the gauge transformations \eqref{general_gaugetransform} read 
\begin{align}
 \label{e:gauge_transf_1} 
 \delta A_\mu^m = & D_\mu \Lambda^m + \frac{1}{2}  \xi^{mn}\, \Xi_{\mu \, n} - \frac{1}{2}  \xi^{ma}\, \Xi_{\mu \, a}\\
 \label{e:gauge_transf_2} \delta A_\mu^a = & D_\mu \Lambda^a +\frac{1}{2}  \xi^{am}\, \Xi_{\mu \, m} -\frac{1}{2}  \xi^{ab}\, \Xi_{\mu \, b}\, . 
\end{align}
The elimination of redundant vectors and tensors is now carried out for the fluctuations around 
the vacuum, rather than at a general point in the unbroken theory.

Recall that there exist orthogonal
matrices $S$, such that the contracted embedding tensors \eqref{e:dressed_gaugings} transform as
\begin{align}\label{e:xi_trafo}
 &S^T \left( \begin{array}{c|c}
 \xi^{mn} &  \xi^{mb} \Bstrut\\ \hline
  \xi^{bn} & \xi^{ab} \Tstrut
                           \end{array}\right) S =
  \left( \begin{array}{c|c}
 \xi^{\hat\alpha \hat\beta} &  \textbf{0}^{\hat\alpha \tilde\beta} \Bstrut \\ \hline
  \textbf{0}^{\tilde\alpha \hat\beta} & \textbf{0}^{\tilde\alpha \tilde\beta} \Tstrut
                           \end{array}\right) \, \\
 &\hat \alpha , \hat \beta =1,\dots , \rk (\xi^{MN}) \, , \qquad \tilde \alpha , \tilde \beta =\rk (\xi^{MN}) +1,\dots , 5+n \, , \nn
\end{align}
where $(\xi^{\hat\alpha \hat\beta})$ is a full-rank matrix.
In particular one can even choose an orthogonal matrix $S$, such that $(\xi^{\hat\alpha \hat\beta})$ is block diagonal
\begin{align}
 (\xi^{\hat\alpha\hat\beta})  
 = \begin{pmatrix}
\gamma_1 \varepsilon &  \cdots & 0 \\ 
\vdots &  \ddots & \vdots \\
 0 & \cdots & \gamma_{n_T} \varepsilon 
                         \end{pmatrix} \, ,
\end{align}
where $n_T = \frac{1}{2} \rk (\xi_{MN})$, which turns out to be the number of complex tensors, and $\varepsilon$ is the two-dimensional epsilon tensor.
The indices $\alpha$, $\hat\alpha$, $\tilde \alpha$ are raised and lowered
with the Kronecker delta.
Along the same lines as before by inverting $(\xi^{\hat\alpha \hat\beta})$ and partial gauge fixing
we find that the propagating degrees of freedom in the vacuum
are captured by $A^{\tilde \alpha}_\mu$ and $B_{\mu\nu\,\hat \alpha}$, where
\begin{align}
&(A^{\alpha}_\mu) =
\left( \begin{array}{c}
 A^{\hat \alpha}_\mu \Bstrut\\
  A^{\tilde \alpha}_\mu 
                           \end{array}\right)
 := S^T \left( \begin{array}{c}
 A^{m}_\mu\\
  A^{a}_\mu 
                           \end{array}\right)\, , 
 & (B_{\mu\nu\, \alpha}) =
 \left( \begin{array}{c}
 B_{\mu\nu\, \hat\alpha} \Bstrut\\
  B_{\mu\nu\, \tilde\alpha}
                           \end{array}\right)
 := S^T \left( \begin{array}{c}
 B_{\mu\nu\, m} \Bstrut\\
  B_{\mu\nu\, a} 
                           \end{array}\right)\, .
\end{align}
The gauge transformations are defined similarly and one easily checks that the fields $A^{\hat \alpha}_\mu$ and $B_{\mu\nu\,\tilde \alpha}$
can be eliminated from the action.
For convenience we also define the dual elements
\begin{align}
& (A^{*\,\alpha}_\mu) =
\left( \begin{array}{c}
 A^{*\,\hat \alpha}_\mu \Bstrut\\
  A^{*\,\tilde \alpha}_\mu 
                           \end{array}\right)
:= S^T \eta S 
 \left( \begin{array}{c}
 \textbf{0}^{\hat \alpha} \Bstrut\\
  A^{\tilde \alpha}_\mu 
                           \end{array}\right) \, , &
 (B^*_{\mu\nu\, \alpha})=
 \left( \begin{array}{c}
 B^*_{\mu\nu\, \hat\alpha}\Bstrut\\
  B^*_{\mu\nu\, \tilde\alpha}
                           \end{array}\right)
 := S^T \eta S 
 \left( \begin{array}{c}
 B_{\mu\nu\, \hat\alpha}\Bstrut\\
  \textbf{0}_{\tilde\alpha} 
                           \end{array}\right)
\end{align}
where $\eta = \textrm{diag}(-1,-1,-1,-1,-1,+1,\dots ,+1)$.
Already at this stage it becomes obvious that the number of complex massive tensors is always given by $\frac{1}{2}\rk ( \xi^{MN} )$.
Moreover a closer look at the Lagrangian \eqref{bos_N=4action} shows that the charge of the tensors is
independent of the vacuum. This will become important in \autoref{sec:effective_action}.
Unfortunately for the vectors such simple statements are not possible, since all properties depend
crucially on the vacuum.

\subsection{The \texorpdfstring{$\cN =4$}{N=4} gauged supergravity action around the vacuum}\label{sec:action}

Applying the redefinition of vectors and tensors of the last section
in order to isolate the propagating degrees of freedom
we are now in a position to derive crucial parts of the action around the vacuum.
In particular, we display
the mass terms and charges of the scalars and tensors, as well as the field strengths,
Chern-Simons terms and mass terms of the vectors in a general form. Inserting the expressions of the
contracted embedding tensors \eqref{e:dressed_gaugings} for a certain example then yields easily the precise spectrum and action.
Furthermore, we derive the formulas for the cosmological constant as well as the gravitino masses.

Before writing down the Lagrangian, let us define the fluctuations of the scalars $\sigma$ and $\cV$ around their VEVs
\begin{align}
 &\sigma = \langle \sigma \rangle + \tilde \sigma \, , \nn \\
 &\cV = \langle \cV \rangle \exp \big(\,\phi^{ma}[t_{ma}]\,\big) \, ,
\end{align}
where $[t_{ma}]_M^{\ N} = 2 \delta_{[m}^{\ \ N} \eta_{a]M}$. The $\phi^{ma}$ capture the unconstrained fluctuations around the VEVs of the coset representatives.
We also define indices $\alpha\, , \beta , \dots$ in expressions like $f_{\alpha m a}$ using the same transformation
as in \eqref{e:xi_trafo}. 
Furthermore, we set
\begin{align}
 \eta_{\alpha\beta} := ( S^T \eta S )_{\alpha\beta} \, ,
\end{align}
where $S$ is the matrix of \eqref{e:xi_trafo} and $\eta = \textrm{diag}(-1,-1,-1,-1,-1,+1,\dots , +1)$.

The relevant part of the Lagrangian of $\cN = 4$ gauged supergravity around the vacuum then reads
\begin{align}\label{e:vac_lagr}
 e^{-1}\cL_{\textrm{rel}} = &  \frac{1}{16 \sqrt 2}\epsilon^{\mu\nu\rho\lambda\sigma}  
  \,\xi^{\hat \alpha \hat \beta}\, B^*_{\mu\nu\, \hat\alpha} \cD_\rho B^*_{\lambda\sigma\, \hat\beta} 
 - \frac{1}{16} \Sigma^2 \, \xi^{\hat\alpha\hat\beta} \tensor{\xi}{_{\hat\alpha}^{\hat\gamma}}\, B^*_{\mu\nu \, \hat\beta} B^{* \,\mu\nu}_{\hat\gamma} \nn \\
 & -\frac{1}{4} \Sigma^2 F^{\tilde \alpha}_{\mu\nu} F_{\tilde \alpha}^{\mu\nu}
 - \frac{1}{4} \Sigma^{-4} F^0_{\mu\nu} F^{0\, \mu\nu}\nn \\
 & - \frac{1}{\sqrt 2} \epsilon^{\mu\nu\rho\lambda\sigma} A_\mu^0 \,\Big ( \partial_\nu A_\rho^{*\, \tilde \alpha} \partial_\lambda A_{\sigma \, \tilde \alpha} 
  - f_{\alpha\beta\gamma}\, A_\nu^{*\, \alpha} A_\rho^{*\, \beta}  \partial_\lambda A_\sigma^{*\, \gamma} 
   - \frac{1}{4} \xi_{\hat\beta\hat\gamma}\, A_\nu^{*\, \hat\beta} A_\rho^{*\, \hat\gamma}  \partial_\lambda A_\sigma^{*\, 0} \Big ) \nn \\
 & - \frac{1}{2} \Big ( \cD_\mu \phi^{ma} - \xi^{ma} A_{\mu}^0 - \tensor{f}{_{\alpha}^m^a} A_{\mu}^{*\, \alpha} \Big ) 
 \Big ( \cD^\mu \phi_{ma} - \xi_{ma} A^{0 \, \mu} - \tensor{f}{^\beta_m_a} A^{*\, \mu}_\beta \Big )\nn \\
& -\frac{1}{2} \partial_\mu \tilde \sigma \, \partial^\mu \tilde \sigma 
  -\frac{1}{2} \textbf{M}^2_{ma\, nb}\, \phi^{ma}\phi^{nb} 
  - \frac{1}{2} \textbf{M}^2 \, \tilde\sigma^2 
  - \textbf{M}^2_{ma} \, \phi^{ma} \tilde \sigma \, ,
\end{align}  
with
\begin{align}
 \cD_\mu \phi^{ma} &:= \partial_\mu \phi^{ma} - A_\mu^0 \, \phi^{nb} \, \big( \tensor{\xi}{_b^a} \delta^m_n -  \tensor{\xi}{_n^m} \delta^a_b \big)
 - A_\mu^{*\, \alpha} \, \phi^{nb} \, \big( \tensor{f}{_\alpha_b^a} \delta^m_n -  \tensor{f}{_\alpha_n^m} \delta^a_b  \big) \, ,\\
\cD_\rho B^*_{\lambda\sigma\, \hat\beta} & :=  \partial_\rho B^*_{\lambda\sigma\, \hat\beta}
 -  \xi^{\hat\gamma\hat\delta}\,\eta_{\hat\beta\hat\delta}\, A^0_\rho \, B^*_{\lambda\sigma\, \hat\gamma} \, \\
 F^{\tilde \alpha}_{\mu\nu} &:= 2\, \partial_{[\mu} A_{\nu]}^{\tilde \alpha} - \tensor{f}{_\beta_\gamma^{\tilde\alpha}}A^{*\, \beta}_\mu A^{*\, \gamma}_\nu \, , \\
 F^0_{\mu\nu} &:= 2\,\partial_{[\mu} A_{\nu]}^{0} \, ,
\end{align}
and
\begin{align}  
\textbf{M}^2_{ma\, nb}:= &  \Sigma^{-2} \Big ( f_{abp}\tensor{f}{_m_n^p} + f_{abc}\tensor{f}{_m_n^c} + f_{anp}\tensor{f}{_m_b^p} + f_{anc}\tensor{f}{_m_b^c}
 + \delta_{mn} f_{acp}\tensor{f}{_b^c^p} + \delta_{ab} f_{mcp}\tensor{f}{_n^c^p} \Big ) \nn \\
& + \frac{1}{3\sqrt 2} \Sigma \Big ( 3\, \varepsilon_{mnpqr}  \tensor{f}{_a_b^p} \xi^{qr}
+ 6\, \varepsilon_{mnpqr}  \tensor{f}{_a^p^q} \tensor{\xi}{_b^r}
 + \varepsilon_{mnpqr} f^{pqr} \xi_{ab} \nn \\
 & +\frac{3}{2} \delta_{ab} \varepsilon_{mspqr}  \tensor{f}{_n^s^p} \xi^{qr}
 - \delta_{ab} \varepsilon_{mspqr} f^{spq} \tensor{\xi}{_n^r} \Big ) \nn \\
& + \frac{1}{2}\Sigma^{4} \Big ( 2\, \xi_{mn} \xi_{ab} + 2\, \xi_{mb} \xi_{an}
 + \delta_{mn} \xi_{ac} \tensor{\xi}{_b^c} + \delta_{mn} \xi_{ap} \tensor{\xi}{_b^p}
   + \delta_{ab} \xi_{mp} \tensor{\xi}{_n^p} + \delta_{ab} \xi_{mc} \tensor{\xi}{_n^c} \Big ) \, , \\
 \textbf{M}^2 :=  & \Sigma^{-2} \Big (- \frac{1}{9} f_{mnp} f^{mnp} + \frac{1}{3} f_{mna} f^{mna} \Big )
 + \frac{4}{3} \,\Sigma^{4}\, \xi^{ma} \xi_{ma} + \frac{1}{18 \sqrt 2} \,\Sigma\, \varepsilon_{mnpqr} f^{mnp} \xi^{qr}\, , \\
\textbf{M}^2_{ma} := &  - \frac{2}{\sqrt 3} \Sigma^{-2} \tensor{f}{_a^b^n} f_{mbn} 
 + \frac{2}{\sqrt 3} \Sigma^4 \Big ( \xi_{ab} \tensor{\xi}{_m^b} + \xi_{an} \tensor{\xi}{_m^n} \Big ) \nn \\
 & + \frac{1}{6 \sqrt 6} \varepsilon_{mnpqr} \Sigma \Big ( 3\, \tensor{f}{_a^n^p} \xi^{qr} -2\, f^{npq} \tensor{\xi}{_a^r} \Big )  \, .
\end{align}
We stress that \eqref{e:vac_lagr} is not the full bosonic Lagrangian around the vacuum, since there are additional couplings
between the fields which are not displayed. 
However, around an $\cN=2$ vacuum the included terms together with the residual supersymmetry
turn out to be sufficient to determine the full effective action, apart from the metric on the quaternionic manifold. In fact, 
as we discuss in more detail for our analysis of the examples in \autoref{sec:ex},   
the effective theory is inferred by knowing the gauge symmetry,
Chern-Simons terms as well as in the masses and charges of the fields.
This data is indeed captured by \eqref{e:vac_lagr}.
It is also important to keep in mind that all contracted embedding tensors are meant to be evaluated in the vacuum.

Let us comment on some of the properties of the action  \eqref{e:vac_lagr}. Closer inspection of \eqref{e:vac_lagr} shows that 
the scalars $\phi^{ma}$ are coupled to the vectors
with standard minimal couplings as well as with St\"uckelberg couplings. This implies that some of the scalars $\phi^{ma}$
constitute the longitudinal degrees of freedom of massive vectors. We also see that it is in general possible to preserve a non-Abelian
gauge group in the vacuum corresponding to a subset of the $A_\mu^{\tilde\alpha}$. For this non-Abelian subgroup 
the corresponding Chern-Simons terms can
in general appear. The tensors are generically charged only under a $U(1)$ gauge symmetry.
As already mentioned, the number of massive tensors is given by $\frac{1}{2}\rk ( \xi^{MN} )$, which is obvious
in \eqref{e:vac_lagr}, since their mass matrix determined by $\xi^{\hat \alpha \hat \beta}$ is full-rank.
 In contrast, note that the mass matrices of vectors and scalars 
are in general not full-rank.

To proceed further one has to specify the precise form of the contracted 
embedding tensors to study the spectrum and action case by case.
In particular, one has to diagonalize the mass matrices or gauge-interaction matrices
of all fields, normalize the kinetic terms, and possibly complexify the fields.
We carry out this procedure for the examples in \autoref{sec:ex}, although not presenting all the details 
of the computations.
The standard Lagrangians of the massive fields are displayed in \autoref{App-Conventions}.
We refer the interested reader to \cite{Grimm:2014soa}, where similar calculations are carried out in detail.

To close this general discussion, let us comment on the cosmological constant in the vacuum. 
The latter can be extracted from the scalar potential in the vacuum, which reads in terms of contracted embedding tensors
\begin{align}\label{e:pot_vac}
 V= - \frac{1}{12}\Sigma^{-2} f^{mnp}f_{mnp} + \frac{1}{4} \Sigma^{-2} f^{mna}f_{mna} + \frac{1}{4}\Sigma^4\, \xi^{am}\xi_{am}
 +\frac{1}{6\sqrt 2}\Sigma\, \varepsilon_{mnpqr} f^{mnp} \xi^{qr}\, .
\end{align}
Furthermore, since we are mainly interested in vacua preserving $\cN=2$ supersymmetry,
it is desirable to have a general expression for the amount of preserved supersymmetry 
for a certain set of contracted embedding tensors.
Since massless gravitinos are in one-to-one correspondence with preserved supersymmetries,
the remaining supersymmetry in the vacuum can be determined from the mass terms of the gravitinos \eqref{fermionic-quadratic}.
The four eigenvalues of the mass matrix $(\tensor{\textbf{M}}{_\psi_\,_i^j})$ are denoted by $\pm m_{\psi\pm}$ given by \cite{Cassani:2012wc}
\begin{align}\label{e:grav_mass}
 m_{\psi\pm} = \sqrt{2\, \textbf{M}_\psi^{mn} \, \textbf{M}_{\psi\, mn} \mp \sqrt{8\, \big(\textbf{M}_\psi^{mn}\, \textbf{M}_{\psi\, mn}\big)^2
 - 16\, \textbf{M}_\psi^{mn}\, \textbf{M}_{\psi\, np}\, \textbf{M}_\psi^{pq}\, \textbf{M}_{\psi\, qm} }}\, .
\end{align}
Additionally the mass of the gravitinos receives contributions from a possibly non-trivial cosmological constant $\Lambda = \langle V \rangle$
\begin{align}\label{e:grav_mass_correction}
 \delta m_\psi = \frac{\sqrt 6}{4} \sqrt{-\langle V \rangle}\, .
\end{align}
The conditions for preserved $\cN=2$ supersymmetry can then be formulated as
\begin{align}
 m_{\psi+} - \delta m_\psi \overset{!}{=} 0 \, .
\end{align}
As explained in \cite{Grimm:2014soa}, for Minkowski vacua, i.e.~$\delta m_\psi=0$, this condition is equivalent to demanding that the eigenvalues
of $\textbf{M}_\psi^{mn}$ coincide in their absolute values.

We have now provided all formulas to check, given a set of contracted embedding tensors, if the associated vacuum preserves supersymmetry and has
a non-trivial cosmological constant. The spectrum and the most relevant terms of the Lagrangian are calculated easily using \eqref{e:vac_lagr}.
In the next section we prepare the application of these results to a class of important examples, namely consistent
truncations of M-theory on $SU(2)$-structure manifolds.

\section{M-theory on \texorpdfstring{$SU(2)$}{SU(2)}-structure manifolds}\label{su2reduction}

In this section we introduce our main examples for a gauged $\cN = 4$ 
supergravity theory in five dimensions by reducing eleven-dimensional 
supergravity on six-dimensional manifolds $\cM_6$ with $SU(2)$-structure.
In \autoref{basicsofSU(2)} we first recall some basic properties 
of $SU(2)$-structure manifolds. The introduced definition will then be used in 
\autoref{sec:reductionansatz} to formulate the reduction ansatz
specifying a consistent truncation of the full compactification on $\cM_6$ to 
five dimensions. The five-dimensional action is derived in \autoref{sec:dimredaction}
and brought into standard $\cN=4$ supergravity form in \autoref{sec:standardN=4form}. 
This allows us to determine the embedding tensors induced by the 
$SU(2)$-structure and a non-trivial flux background.

\subsection{Some basics on \texorpdfstring{$SU(2)$}{SU(2)}-structure manifolds}\label{basicsofSU(2)}

Let us begin by recalling some basics on six-dimensional $SU(2)$-structure manifolds $\cM_6$.
See e.g. \cite{Hitchin:2000sk, hitchin2001stable, chiossi2002intrinsic, Gauntlett:2002sc, Gauntlett:2003cy} for properties of general \(G\)-structure manifolds and \cite{Bovy:2005qq, ReidEdwards:2008rd, Lust:2009zb, Triendl:2009ap, Louis:2009dq, Danckaert:2011ju, KashaniPoor:2013en} for \(SU(2)\)-structure manifolds.
If the structure group of a manifold $\cM_6$ can be reduced to \(SU(2)\), it admits two globally defined, 
nowhere vanishing spinors \(\eta^1\), \(\eta^2\). This can be seen from the decomposition of the spinor representation ${\bf 4} $ of $Spin(6) \cong SU(4)$ into \(SU(2)\) representations,  \({\bf 4} \rightarrow 2 \cdot {\bf 1} \oplus {\bf 2}\).
The existence of these two spinors gives rise to four supersymmetry generators \(\xi^{1,2}_i\) (\(i = 1,2\)) in five dimensions, since we can write 
the eleven-dimensional supersymmetry generator \(\epsilon\) as 
\begin{equation}
\epsilon = \xi^1_i \otimes \eta^i + \xi^2_i \otimes \eta^{c\, i} \,,
\end{equation}
where $\eta^{c\, i}$ is the charge conjugate spinor to $\eta^i$ and the five-dimensional spinors \(\xi^{1,2}_i\) are symplectic Majorana, see
\autoref{App-Conventions}.
This implies that 
an appropriately chosen reduction
admits $\cN=4$ supersymmetry. 

The globally defined spinors $ \eta^i$ allow to define  three real two-forms \(J^a\), \(a = 1, 2, 3\), forming a \(SU(2)\) triplet,
and a complex one-form $K$. These fulfill the conditions
\begin{equation}\begin{gathered}\label{eq:su2forms}
J^a \wedge J^b = \delta^{ab} \mathrm{vol}_4 \,, \\
K_m K^m = 0 \,,\quad
\bar K_m K^m = 2\ ,\quad
K^m J^a_{mn} = 0 \,,
\end{gathered}\end{equation}
where \(m,n = 1, \dots, 6\) and $\mathrm{vol}_4$ is a no-where vanishing four-form on $\cM_6$.
All contractions are performed with the $SU(2)$-structure metric on $\cM_6$. 

These forms define an almost product structure
\begin{equation}
{P_m}^n = K_m \bar K^n + \bar K_m K^n - {\delta_m}^n \,, 
\end{equation}
which allows to split the manifold's tangent space into the eigenspaces of \(P\),
\begin{equation} \label{tangetsplit}
T\cM_6 = T_2 \cM_6 \oplus T_4 \cM_6 \,,
\end{equation}
where the part $ T_2 \cM_6$ is spanned by $K^1 = \R\, K$ and $K^2 = \I\, K$.

\subsection{The reduction ansatz}\label{sec:reductionansatz}

An appropriate ansatz for the dimensional reduction on manifolds with structure group \(SU(2)\) 
has been worked out in \cite{Danckaert:2011ju,KashaniPoor:2013en}.
The full spectrum of the compactified theory consists of infinitely many modes from which the choice of a particular ansatz keeps only a finite subset.
Such a truncation is called consistent, if any of the modes that we keep cannot excite one of modes we exclude.
This means that there are no source terms for the discarded fields in the reduced action.
In this case any solution of the truncated theory can be uplifted to a solution of the full eleven-dimensional equations of motion.
As explained in \cite{KashaniPoor:2013en} this can be achieved by choosing the reduction ansatz to be a set of forms on \(\cM_6\) that it is closed under the action of the wedge product \(\wedge\), exterior differentiation \(\dd\) and the Hodge star \(\ast\).

In \cite{Triendl:2009ap} it has been demonstrated how to decompose the field content of type IIA supergravity into representations with respect to the \(SU(2)\) structure group of \(\cM_6\) and arrange it into four-dimensional \(\cN = 4\) multiplets.
The same analysis can be performed for the case of eleven-dimensional supergravity reduced to \(\cN = 4\) supergravity in five dimensions.
The modes transforming as singlets under \(SU(2)\) constitute the five-dimensional gravity multiplet and a pair of vector multiplets,
and every \(SU(2)\)-triplet corresponds to one triplet of vector multiplets.
On the other hand the components of the fields that are doublets under \(SU(2)\) form gravitino multiplets in the \(\cN = 4\) theory.
Since it is not known how to consistently couple gravitino multiplets to gauged \(\cN = 4\) supergravity, these multiplets will be neglected.
This is equivalent to excluding all \(SU(2)\) doublets from the reduction ansatz.
We will further comment on this point in \autoref{sec:ex1}.

Following up these considerations the reduction ansatz consists now of a basis of real one-forms \(v^i\) (\(i = 1,2\)) on \(T_2\cM_6\),
and real two-forms \(\omega^I\) (\(I = 1, \dots, \tilde n\)) on \(T_4\cM_6\). Forms of odd rank on \(T_4\cM_6\) correspond to doublets of \(SU(2)\) and are thus not included in the ansatz.
These forms are normalized via
\begin{equation}\label{eq:su2ansatzintegral}
\int_{\cM_6} v^1 \wedge v^2 \wedge \omega^I \wedge \omega^J = -\eta^{IJ} \,,
\end{equation}
where \(\eta^{IJ}\) is an \(SO(3,\tilde n-3)\) metric that will be used to raise and lower indices.
For convenience we can also introduce \(\mathrm{vol}_2^{(0)} = v^1 \wedge v^2\) and \(-\eta^{IJ} \mathrm{vol}_4^{(0)} = \omega^I \wedge \omega^J\), which take the role of normalized volume forms on \(T_2\cM_6\) and \(T_4\cM_6\) respectively.

The ansatz has to be chosen such that it is consistent with 
exterior differentiation. Therefore, we demand that the differentials of \(v^i\) and \(\omega^I\) obey 
\begin{equation}\begin{aligned}\label{eq:torsion}
\dd v^i &= t^i \,v^1 \wedge v^2 + t^i_I \,\omega^I\,, \\
\dd \omega^I &= T^I_{iJ} \,v^i \wedge \omega^J\,,
\end{aligned}\end{equation}
where the coefficients \(t^i\), \(t^i_I\) and \(T^I_{iJ}\) are related to the torsion classes of $\cM_6$
and have to fulfill the consistency conditions \cite{KashaniPoor:2013en}
\begin{equation}\begin{aligned}\label{eq:torsionconditions}
t^i t^k_I \epsilon_{kj} + t^i_J T^J_{jI} = 0 \,&,\quad
T^I_{iJ} \eta^{JK} t^i_K = 0 \,,\\
T^I_{iJ} t^i - T^I_{iK} \epsilon_{ij} T^K_{jJ} = 0 \,&,\quad
t^i \eta^{IJ} - \epsilon_{ij} T^I_{jK} \eta^{KJ} - \epsilon_{ij} T^J_{jK} \eta^{KI} = 0 \,.
\end{aligned}\end{equation}
Using this basis of forms, one now has to expand 
all fields of eleven-dimensional supergravity. In order to 
discuss the reduction of the eleven-dimensional action, 
we first expand the $J^a$ and $K$ introduced in \eqref{eq:su2forms}
as
\begin{equation}\label{eq:su2formsexpansion}
J^a = e^{\rho_4/2} \zeta^a_I \omega^I\ ,\qquad 
K = e^{\rho_2/2} (\I\,\tau)^{-1/2} (v^1 + \tau v^2) \,,
\end{equation}
where now the real $\rho_4$, $\rho_2$, $\zeta^a_I$, and complex $\tau$ 
are promoted to five-dimensional space-time scalars.
Together with \eqref{eq:su2forms} we find \(\zeta^a_I \eta^{IJ} \zeta^b_J = -\delta^{ab}\) as well as \(\mathrm{vol}_4 = e^{\rho_4} \mathrm{vol}_4^{(0)}\) and \(K^1 \wedge K^2 = e^{\rho_2} \mathrm{vol}_2^{(0)}\).

The action of the Hodge star on the ansatz is given by
\begin{equation}\begin{aligned}\label{eq:su2ansatzhodge}
\ast\, v^i &= e^{\rho_4}\, \epsilon_{ij} v^j \wedge \mathrm{vol}_4^{(0)} \,, \\
\ast\, \mathrm{vol}_2^{(0)} &= e^{\rho_4}\, \mathrm{vol}_4^{(0)} \,, \\
\ast\, \omega^I &= - e^{\rho_2} {H^I}_J \omega^J \wedge \mathrm{vol}_2^{(0)} \,, \\
\ast \left(v^i \wedge \omega^I\right) &= - \epsilon_{ij} {H^I}_J v^j \wedge \omega^J \,.
\end{aligned}\end{equation}
From the requirement that \(\ast J^a = J^a \wedge K^1 \wedge K^2\) the matrix \({H^I}_J\) can be determined to be \(H_{IJ} = 2\zeta^a_I\zeta^a_J + \eta_{IJ}\).
See \autoref{app:cosetrepr} for a further discussion of its properties.

After this preliminary discussion we are now in a position to give the ansatz for the eleven-dimensional metric. 
More precisely, reflecting the split of the tangent space \eqref{tangetsplit}, the metric takes the 
form 
\begin{equation}\label{e:su(2)_metric}
\dd s^2_{11} = g_{\mu\nu} \dd x^\mu \dd x^\nu + e^{\rho_2}g_{ij} (v^i + G^i)(v^j + G^j) + e^{\rho_4} g_{st} \dd x^s \dd x^t \,,
\end{equation}
with \(s,t=1,\dots,4\).
 The \(G^i\) are space-time gauge fields parameterizing the variation of $T_2 \cM_6$.
The metric \(g_{ij}\) can be expressed in terms of \(\tau\),
\begin{equation}
g = \frac{1}{\I\,\tau} \begin{pmatrix}1 & \R\,\tau \\ \R\,\tau & |\tau|^2 \end{pmatrix} \ ,
\end{equation}
such that $e^{\rho_2} g_{ij} v^i v^j = K \bar K$.
Notice that we excluded possible off-diagonal terms of the form \(g_{\mu s}\) and \(g_{is}\) from the ansatz for the metric that would precisely correspond to \(SU(2)\) doublets.
These terms would give rise to two doublets of additional space-time vectors and four doublets of space-time scalars.

In the following it will be useful to introduce the 
gauge invariant combination
\begin{equation}
\tiv^i = v^i + G^i \,.
\end{equation}
whose derivative can be calculated using \eqref{eq:torsion},
\begin{equation}
\dd \tiv^i = \dd \!\left(v^i + G^i\right) = \D G^i + t^i \tiv^1 \wedge \tiv^2 - t^i \epsilon_{jk} \tiv^j \wedge G^k + t^i_I \omega^I \,.
\end{equation}
The definition of the covariant derivative \(\D G^i\) can be found in \eqref{eq:covderivvectors}.

Let us next turn to the ansatz for the three form field \(C_3\). 
Using the basis $\tilde v^i,\omega^I$ introduced above, we expand
\begin{equation}\label{eq:c3expansion}
C_3 = C + C_i \wedge \tiv^i + C_I \wedge \omega^I + C_{12} \wedge \tiv^1 \wedge \tiv^2 + c_{iI} \, \tiv^i \wedge \omega^I\,.
\end{equation}
If we had included \(SU(2)\) doublets in the reduction ansatz, we would have also had to expand \(C_3\) in terms of odd forms on \(T_4\cM_6\) \footnote{To make the ansatz closed under wedge product it might be necessary in this case to include also additional two-forms on \(T_4\cM_6\) and hence additional \(SU(2)\) triplets.},
which would give additional fields in five dimensions.
For each doublet these would be one doublet of two-forms and two doublets of vectors and scalars.
Together with the contributions from the metric, we see that for every excluded \(SU(2)\) doublet this resembles precisely a doublet of \(\cN = 4\) gravitino multiplets.

Furthermore, we consider also a possible internal four-form flux for which the most general ansatz is given by
\begin{equation}
F_4^\mathrm{flux} = n\, \mathrm{vol}_4^{(0)} + n_I\, v^1 \wedge v^2 \wedge \omega^I \,.
\end{equation}
Notice that this is written only in terms of \(v^i\) and not in terms of the gauge invariant quantities \(\tiv^i\), because this would
introduce an unwanted space-time dependency. Moreover \(n\) and \(n_I\) are not completely independent,
since it follows from \(\dd F_4^\mathrm{flux} = 0\) that
\begin{equation}\label{eq:fluxcondition}
n \,t^i - n^I t^i_I = 0 \,.
\end{equation}
We finally have to expand the field strength \(F_4 = F_4^\mathrm{flux} + \dd C_3\), 
\begin{equation}\begin{aligned}\label{eq:f4expansion}
F_4 &=  F + F_i \wedge \tiv^i + F_I \wedge \omega^I + F_{12} \wedge \tiv^1 \wedge \tiv^2 + F_{iI} \, \tiv^i \wedge \omega^I \\
&\quad + f_I \, \tiv^1 \wedge \tiv^2 \wedge \omega^I + f \vol_4^{(0)} \,,
\end{aligned}\end{equation}
and obtain after calculating the derivative of \(C_3\) the expansion coefficients
\begin{equation}\begin{aligned}\label{eq:fieldstrengths}
F &= \dd C + C_i \wedge \D G^i \,, \\
F_i &= \D C_i + \epsilon_{ij} C_{12} \wedge \D G^j \,, \\
F_I &= \D C_I + c_{iI} \D G^i \,, \\
F_{12} &= \D C_{12} \,, \; F_{iI} = \D c_{iI} \,, \\
f_I &= n_I + t^i c_{iI} + \epsilon_{ij} T^J_{iI} c_{jJ} \,, \\
f &= n-c_{iI}t^i_J \eta^{IJ} \,. 
\end{aligned}\end{equation}
The four-form flux  and the fact that \(\omega^I\) and \(\tiv^i\) are in general non-closed forms 
induce various non-trivial gaugings. These are encoded by the various appearing covariant derivatives that 
are listed in the next subsection.

\subsection{Dimensional reduction of the action} \label{sec:dimredaction}

Starting from the bosonic action of eleven-dimensional supergravity,
\begin{equation}\label{eq:mtheoryaction}
S = \int_{11} \tfrac{1}{2} (\ast 1) R - \tfrac{1}{4} F_4 \wedge \ast F_4 - \tfrac{1}{12}C_3 \wedge F_4 \wedge F_4\,,
\end{equation}
we will compute a five dimensional action by compactifying it on \(\cM_6\). We can compare the result with the general description of \(\cN = 4\)
gauged supergravity given above and determine the embedding tensors in terms of geometrical properties of \(\cM_6\).

To compute the reduced five-dimensional action we insert the expansions \eqref{eq:c3expansion} and \eqref{eq:f4expansion} into the eleven-dimensional action \eqref{eq:mtheoryaction} and integrate over the internal manifold using \eqref{eq:su2ansatzintegral}.
The reduction of the Einstein-Hilbert term has been done in \cite{KashaniPoor:2013en} and can be adopted without further modifications.
After performing an appropriate Weyl rescaling \(g_{\mu\nu} \rightarrow e^{-\frac{2}{3}(\rho_2 + \rho_4)} g_{\mu\nu}\), to bring the action into the Einstein frame, the final result reads 
\begin{equation}\begin{aligned}\label{eq:su2action}
S_{SU(2)} &= \int_5 \biggl\{\tfrac{1}{2} (\ast 1) R_5
- e^{\frac{5}{3}\rho_2 + \frac{2}{3}\rho_4}g_{ij} \D G^i \wedge \ast \D G^j 
-\tfrac{1}{2} (\eta^{IJ} + \zeta^{bI}\zeta^{bJ})\D\zeta^a_I \wedge \ast \D\zeta^a_J\\
& -\tfrac{1}{4}(\I\,\tau)^{-2}\,\D\tau\wedge\ast\D\bar\tau
- \tfrac{5}{12} \D\rho_2 \wedge \ast \D\rho_2
- \tfrac{1}{3}\D\rho_2 \wedge \ast \D\rho_4
-\tfrac{7}{24}\D\rho_4\wedge\ast\D\rho_4 \\
& -\tfrac{1}{4} e^{2(\rho_2+\rho_4)} \left(\dd C + C_i \wedge \D G^i\right) \wedge \ast \left(\dd C + C_j \wedge \D G^j \right) \\
& -\tfrac{1}{4} e^{\frac{1}{3}\rho_2+\frac{4}{3}\rho_4} \left(g^{-1}\right)^{ij} \left(\D C_i + \epsilon_{ik} C_{12} \wedge \D G^k\right) \wedge \ast \left(\D C_j + \epsilon_{jl} C_{12} \wedge \D G^l\right) \\
& -\tfrac{1}{4} e^{\frac{2}{3}\rho_2-\frac{1}{3}\rho_4} H^{IJ} \left(\D C_I + c_{iI} \D G^i\right) \wedge \ast \left(\D C_J + c_{jJ} \D G^j\right) \\
& -\tfrac{1}{4} e^{-\frac{4}{3}\rho_2+\frac{2}{3}\rho_4} \D C_{12} \wedge \ast \D C_{12} - \tfrac{1}{4} e^{-\rho_2-\rho_4} H^{IJ} \left(g^{-1}\right)^{ij} \D c_{iI} \wedge \ast \D c_{jJ} \\
& + \left(\tfrac{1}{4} \dd C + \tfrac{1}{6} C_k \wedge \D G^k\right)\wedge c_{iI}\left(\epsilon^{ij}T^K_{jJ} C_K  + C_{12}t^i_J + \D c_{jJ}\epsilon^{ij}\right) \eta^{IJ} \\
& - \tfrac{1}{6}C_i \wedge \epsilon^{ij}\Bigl( \left(\D C_j + \epsilon_{jk} C_{12} \wedge \D G^k\right) c_{kI}t^k_J + \left(\D C_I + c_{kI} \D G^k\right) \wedge \D c_{jJ}\Bigr)\eta^{IJ} \\
& + \tfrac{1}{6}C_I \wedge \Bigl(\left(\D C_i + \epsilon_{ik} C_{12} \wedge \D G^k\right) \wedge \D c_{jJ}\epsilon^{ij} + \left(\D C_J + c_{lJ} \D G^l\right) \wedge \D C_{12}\Bigr)\eta^{IJ} \\
&+ \tfrac{1}{12} C_{12} \wedge \left(\D C_I + c_{iI} \D G^i\right) \wedge \left(\D C_J + c_{jJ} \D G^j\right)\eta^{IJ} \\
& - \tfrac{1}{6}c_{iI}  \left(\D C_j + \epsilon_{jk} C_{12} \wedge \D G^k\right) \wedge \left(\D C_J + c_{lJ} \D G^l\right) \epsilon^{ij} \eta^{IJ} \\
& - \tfrac{1}{4} n\,\epsilon^{ij} C_i \wedge \left(\D C_i + \epsilon_{ik} C_{12} \wedge \D G^k\right)
 - \left(\tfrac{1}{2} \dd C + \tfrac{1}{4} C_i \wedge \D G^i\right) \wedge \left(n \, C_{12} - n^I C_I\right) \\
& + (\ast 1) \,V \biggl\}\,.
\end{aligned}\end{equation}
The potential term \(V\) is given by
\begin{equation}\begin{split}
V = -\tfrac{5}{8} e^{-\frac{5}{3}\rho_2 - \frac{2}{3}\rho_4} g_{ij} t^i t^j
+ 2 e^{\frac{1}{3}\rho_2 - \frac{5}{3}\rho_4}  g_{ij} t^i_I t^j_J \eta^{IJ} \\
- \tfrac{1}{2} e^{-\frac{5}{3}\rho_2 - \frac{2}{3}\rho_4} (\eta^{IJ} + \zeta^{bI}\zeta^{bJ})\zeta^a_K\zeta^a_L g^{ij} \tilde{T}^K_{iI} \tilde{T}^L_{jJ} \\
+ \tfrac{1}{4}e^{-\frac{8}{3}\rho_2-\frac{5}{3}\rho_4} H^{IJ} f_I f_J
+ \tfrac{1}{4}e^{-\frac{2}{3}\rho_2-\frac{8}{3}\rho_4} f^2 \,.
\end{split}\end{equation}
As mentioned above we have defined several covariant derivatives.
For the scalars they are given by
\begin{equation}\begin{aligned}\label{eq:covderivscalars}
\D\rho_2 &= \dd\rho_2 - \epsilon_{ij} G^i t^j \,, \\
\D\rho_2 &= \dd\rho_4 + \epsilon_{ij} G^i t^j \,, \\
\D\tau &= \dd\tau - ((1,\tau)\cdot G)((1, \tau) \cdot t) \,, \\
\D\zeta^a_I &= \dd\zeta^a_I - G^i\tilde{T}^J_{iI}\zeta^a_J \,, \\
\D c_{iI} &= \dd c_{iI} + \epsilon_{ij} t^j_I C_{12} - T^J_{iI} C_J + \epsilon_{ij} G^j t^k c_{kI} - G^j T^J_{jI} c_{iJ} + n_I \epsilon_{ij} G^j \,,
\end{aligned}\end{equation}
whereas those of the vectors read
\begin{equation}\begin{aligned}\label{eq:covderivvectors}
\D G^i &= \dd G^i - t^i G^1 \wedge G^2\,, \\
\D C_I &= \dd C_I + t^i_I C_i + T^J_{iI} C_J \wedge G^i - n_I G^1 \wedge G^2 \,, \\
\D C_{12} &= \dd C_{12} + t^i C_i  - \epsilon_{ij}C_{12} \wedge t^i G^j \,. \\
\end{aligned}\end{equation}
There is also a pair of two-forms \(C_i\) with
\begin{equation}
\D C_i = \dd C_i + \epsilon_{ij} G^j \wedge t^k C_k \,.
\end{equation}

In the next subsection we compare \eqref{eq:su2action} with the general form of gauged \(\cN = 4\) supergravity.
For this purpose it is necessary to dualize the three-form field \(C\) into a scalar \(\gamma\).
Let us therefore collect all terms from the action containing it,
\begin{equation}\begin{aligned}\label{eq:caction}
S_C = &\int -\frac{1}{4}e^{2(\rho_2+\rho_4)} F \wedge \ast F + \frac{1}{2} F \wedge L \,,
\end{aligned}\end{equation}
with 
\begin{equation}
L = \tfrac{1}{2}c_{iI}\left(\epsilon^{ij}T^K_{jJ} C_K  + C_{12}t^i_J + \D c_{jJ}\epsilon^{ij}\right)\eta^{IJ} - n\, C_{12} + n^I C_I \,.
\end{equation}
The field strength \(F = \dd C + C_i \wedge \D G^i\) fulfills the Bianchi identity
\begin{equation}
\dd F = \D C_i \wedge \D G^i \,,
\end{equation}
which we will impose by introducing a Lagrange multiplier \(\gamma\). Accordingly we add the following term to the action
\begin{equation}
\delta S = -\frac{1}{2} \int \gamma \left(\dd F - \D C_i \wedge \D G^i\right) \,.
\end{equation}
We can now use the equation of motion for \(F\),
\begin{equation}
- e^{2(\rho_2+\rho_4)} \ast F + L + \dd\gamma = 0
\end{equation}
to eliminate it from \eqref{eq:caction} and obtain
\begin{equation}\begin{aligned}
S_{\gamma} = &- \frac{1}{4} \int e^{-2(\rho_2+\rho_4)} (\D\gamma + \tfrac{1}{2}c_{iI} \D c_{jJ}\epsilon^{ij} \eta^{IJ}) \wedge \ast (\D\gamma + \tfrac{1}{2}c_{iI} \D c_{jJ}\epsilon^{ij} \eta^{IJ}) \\
&+ \frac{1}{2} \int \gamma \D C_i \wedge \D G^i \,,
\end{aligned}\end{equation}
where the covariant derivative of \(\gamma\) is defined as
\begin{equation}\label{eq:covderivgamma}
\D\gamma = \dd\gamma + \tfrac{1}{2}c_{iI}(\epsilon_{ij}T^K_{jJ} C_K + t^i_J C_{12})\eta^{IJ} - n\, C_{12} + n^I C_I \,.
\end{equation}

Moreover in the general \(\cN = 4\) theory there are no tensors with second order kinetic term, so it is necessary to trade the two-form \(C_i\) for its dual vector \(\tilde C^{\ib}\).
But since \(C_i\) appears additionally in the covariant derivatives of the vectors \(C_I\) and \(C_{12}\), it will be necessary to introduce their duals \(\tilde{C}_I\) and \(\tilde{C}_{12}\) as well.
These dualizations are described for the case of type IIA supergravity reduction  in \cite{Danckaert:2011ju} and \cite{KashaniPoor:2013en}, so we will not
perform the explicit calculations again.

\subsection{Comparison with \texorpdfstring{$\cN = 4$}{N=4} supergravity} \label{sec:standardN=4form}

As we have described above, the reduced action possesses \(\cN = 4\) supersymmetry, so we will work out how to identify it with the general description of gauged \(\cN = 4\) supergravity from \autoref{N=4Gen}.

The arrangement of the vectors into \(SO(5,n)\) representations \(A^M\) and \(A^0\) and the form of the scalar metric \(M_{MN}\) can be worked out easiest by switching off all gaugings, i.e.~setting \(t^i = t^i_I = T^I_{iJ} = 0\) and \(n = n_I = 0\).
Since in this way all covariant derivatives become trivial and some of the terms in \eqref{eq:su2action} vanish, it is now very easy to carry out
the dualization of \(C^i\) explicitly.
Afterwards the theory will contain \(5 + \tilde n\) vectors in total, which means that there are \(\tilde n - 1\) vector multiplets and the global symmetry group is given by \(SO(1,1) \times SO(5,\tilde n-1)\).
It is natural to identify \(C_{12}\), which does not carry any indices, with the \(SO(5,\tilde n-1)\) scalar \(A^0\) and the other vectors with \(A^M\), so in summary we have
\begin{equation}\begin{aligned}\label{eq:su2vectors}
A^M &= \left(G^i, \tilde{C}^{\bar\imath}, C^J \right)\,, \\ 
A^0 &= C_{12}\,.
\end{aligned}\end{equation}
The corresponding \(SO(5,\tilde n-1)\) metric is defined as\footnote{Note that in the standard form of gauged supergravity $\eta$ is taken
to be diagonal. Therefore,
in order to compare fields and embedding tensors in this reduction to their standard form, one has to diagonalize $\eta$, which is easily done.}
\begin{equation}\label{eq:su2metric}
\eta_{MN} =
\begin{pmatrix}
0 & \delta_{i\bar\jmath} & 0 \\
\delta_{i\bar\jmath} & 0 & 0 \\
0 & 0 & \eta_{IJ}
\end{pmatrix}\,.
\end{equation}
By comparing the kinetic terms of the vectors (in the ungauged theory) with \eqref{bos_N=4action} one obtains the scalar
\begin{equation}\label{eq:sigma}
\Sigma = e^{\frac{1}{3}\rho_2-\frac{1}{6}\rho_4}\,,
\end{equation}
and the coset metric
\begin{equation}\begin{aligned}\label{eq:scalarmetric}
M_{ij} &= e^{\rho_2 + \rho_4} g_{ij} + H_{IJ}\, c^I_i c^J_j + e^{-\rho_2-\rho_4}g^{kl}(\epsilon_{ki}\gamma + \tfrac{1}{2} c_{kI} c^I_i)(\epsilon_{lj}\gamma + \tfrac{1}{2} c_{lI} c^I_j)\,, \\
M_{i\jb} &= e^{-\rho_2-\rho_4} g^{jk}\delta_{j\jb} (\epsilon_{ki}\gamma + \tfrac{1}{2} c_{kI} c^I_i)\,, \\
M_{iI} &= -H_{IJ} c^J_i + e^{-\rho_2-\rho_4} g^{jk} c_{jI} (\epsilon_{ki}\gamma + \tfrac{1}{2} c_{kI} c^I_i)\,, \\
M_{\ib\jb} &= e^{-\rho_2-\rho_4}g^{ij}\delta_{i\ib}\delta_{j\jb}\,, \\
M_{\ib I} &= e^{-\rho_2-\rho_4}g^{ij}\delta_{i\ib}c_{jI}\,, \\
M_{IJ} &= H_{IJ} + e^{-\rho_2-\rho_4}g^{ij}c_{iI}c_{jJ}\,.  
\end{aligned}\end{equation}
From this metric one can also determine the coset representative \(\cV = ({\cV_M}^m, {\cV_N}^a)\), where \(m\) and \(a\) are \(SO(5)\) or \(SO(\tilde n-1)\) indices respectively. \(\cV\) is related to the scalar metric via \(M = \cV\cV^T\) and carries the same amount of information. The result can be found in \autoref{app:cosetrepr}.

From \eqref{eq:sigma} and \eqref{eq:scalarmetric} we can calculate the general covariant derivatives of the scalars using \eqref{gen_cov_der} and compare them with the results from \eqref{eq:covderivscalars} and \eqref{eq:covderivgamma} to derive the embedding tensors
\begin{equation}\begin{aligned}\label{eq:su2embeddingtensors}
\xi_i &= - \epsilon_{ij}t^j\,, \\
\xi_{iI} &= \epsilon_{ij}t^j_I\,, \\
f_{ij\bar\imath} &= \delta_{\bar\imath[i}\epsilon_{j]k}t^k\,, \\
f_{iIJ} &= - T^K_{iI} \eta_{KJ} - \tfrac{1}{2}\epsilon_{ij}t^j \eta_{IJ} \,,
\end{aligned}\end{equation}
and
\begin{equation}\begin{aligned}\label{eq:su2embeddingtensorsflux}
\xi_{ij} &= \epsilon_{ij} n \,, \\
f_{ijI} &= - \epsilon_{ij} n_I \,. \\
\end{aligned}\end{equation}
All other components are either determined by antisymmetry or vanish. One can now use these expressions to calculate the covariant derivatives of the vectors from \eqref{gen_cov_der} and check that they agree with \eqref{eq:covderivvectors}.

To show that the quadratic constraints on the embedding tensors from \eqref{eq:su2embeddingtensors} hold, it is necessary to use the consistency relations \eqref{eq:torsionconditions} on the matrices \(t^i\), \(t^i_I\) and \(T^I_{iJ}\), while the quadratic constraints involving \eqref{eq:su2embeddingtensorsflux} are fulfilled due to \eqref{eq:fluxcondition}.

If we neglect the contributions coming from the four-form flux, it is possible to check that \eqref{eq:su2embeddingtensors} is consistent with the results from the type IIA reduction in \cite{KashaniPoor:2013en}.
This is described in \autoref{IIAreduction}.

\section{Partial supergravity breaking applied to consistent truncations}\label{sec:ex}

In this section we elaborate on the general discussion 
of supersymmetry breaking in \autoref{N=4Generalities} by investigating 
concrete examples given by consistent truncations of higher dimensional theories.
In particular we analyze their quantum effective action. In \autoref{sec:effective_action}
we start with general considerations on the effective action of consistent truncations.
The analysis of one-loop Chern-Simons terms allows us to formulate necessary conditions such that
a consistent truncation gives rise to a physical sensible effective theory.
One class of examples, worked out in \autoref{sec:ex1}, will be provided by the $SU(2)$-structure 
reductions of \autoref{su2reduction} with Calabi-Yau vacuum.  Closely related to these kind of reductions is a second class of examples,
consistent truncations of type IIB supergravity on squashed Sasaki-Einstein manifolds, which we investigate in \autoref{sec:ex2}.

\subsection{Quantum effective action of consistent truncations}\label{sec:effective_action}

We start by studying the quantum effective action 
obtained after $\cN=4 \rightarrow \cN=2$ spontaneous supersymmetry 
breaking. An effective action is obtained by fixing a certain energy scale
and integrating out all modes that are heavier than this scale. In 
five dimensions this is particularly interesting, since massive charged modes
induce Chern-Simons terms at one-loop. Importantly, these corrections
do not dependent on the masses of the modes in the loop and are therefore never suppressed.
We are interested in evaluating these terms for
the supersymmetry breaking mechanism in \autoref{N=4Generalities}.
A prominent class of examples for this pattern
is given by consistent truncations of supergravity.
For instance, if a Calabi-Yau manifold has $SU(2)$-structure, the $\cN=4$ gauged supergravity from the M-theory reduction in the previous
section is broken to $\cN=2$ in the vacuum.
It is an interesting question when a consistent truncation also gives rise to a 
proper effective theory. For example, in order to phenomenologically analyze 
non-Calabi-Yau reductions of string theory or M-theory one needs to deal with 
effective theories. 
The fact that Chern-Simons terms are independent of the mass scale allows
us thus to investigate the question:
\begin{itemize}
 \item  What are the necessary conditions for a consistent truncation to yield 
 the physical effective theory of the setup below a cut-off scale where all massive 
 modes are integrated out?
\end{itemize}
Indeed the one-loop Chern-Simons terms of the consistent truncation should match the ones of the genuine
effective action. Clearly a first step is to analyze compactifications of which we know the relevant parts of the effective theory,
like Calabi-Yau compactifications.

It was found in \cite{Bonetti:2013ela,Bonetti:2012fn} that in five dimensions Chern-Simons terms can receive corrections from
massive charged fermions and self-dual tensors at one-loop.
The classical gauge and gravitational Chern-Simons terms
\begin{align}\label{e:standard_CS}
 e^{-1} \cL_{CS} =  \frac{1}{48} \epsilon^{\mu\nu\rho\sigma\tau} 
 \, k_{I J K} \, A^I_\mu F^J_{\nu\rho} F^K_{\sigma\tau} 
 + \frac{1}{16} \epsilon^{\mu\nu\rho\sigma\tau} 
 \, k_{I} \, A^I_\mu \tensor{R}{^a_b_\nu_\rho} \tensor{R}{^b_a_\sigma_\tau} 
\end{align}
($\tensor{R}{^a_b_\mu_\nu}$ are the components of the curvature two-form)
get corrected by a shift for each massive charged fermion and tensor mode
\begin{align}
 k_{\Lambda\Sigma\Theta} &\mapsto k_{\Lambda\Sigma\Theta} + c_{AFF}\, q_\Lambda q_\Sigma q_\Theta\, \sign (\cR) \\
 k_{\Lambda} &\mapsto k_{\Lambda} + c_{A\cR\cR}\, q_\Lambda\, \sign (\cR)  \, ,
\end{align}
where the constants $c_{AFF}$, $c_{A\cR\cR}$ are given in \autoref{t:CS_correct} and the sign of the representation is defined as
\begin{align}
 \sign (\cR)= \begin{cases} 
               +1\, , & \textrm{for } \cR = (\frac{1}{2},0) , (1,0) , (\frac{1}{2},1) \\
               -1\, , & \textrm{for } \cR = (0,\frac{1}{2}) , (0,1) , (1,\frac{1}{2}) \, ,
              \end{cases}
\end{align}
and the representations $\cR$ are labeled by their spin with respect to $SU(2)\times SU(2)\cong SO(4)$, the massive little group.
\begin{table}
\begin{center}
\begin{tabular}{llll}
\hline\hline
 & spin-1/2 fermion & self-dual tensor & spin-3/2 fermion\\
 \hline
$c_{AFF}$ & $1/2$ & $-2$ & $5/2$\\
$c_{A\cR\cR}$ & $-1$ & $-8$ & $19$\\
\hline\hline
\end{tabular}
\end{center}
\caption{One-loop Chern-Simons coefficients}
\label{t:CS_correct}
\end{table}
As one can see, the contributions are independent of the mass scale and indeed they are related to anomalies.
Because of this property they capture crucial information about the massive modes. For example in F-theory compactifications
they often suffice to calculate the whole spectrum \cite{Bonetti:2011mw,Grimm:2013oga}.
In this spirit we think that one-loop corrections to Chern-Simons terms can teach us lessons about the question
when consistent truncations also yield proper effective field theories.

In particular a necessary condition for such a reduction to make sense as an effective field theory after integrating out massive modes is that
the one-loop Chern-Simons terms should coincide with the ones in the genuine effective action. Stated differently, the corrections to the Chern-Simons terms
induced by the truncated modes must coincide with the ones which are obtained by taking the full infinite tower of massive modes into account.
For the special case that the relevant parts of the effective theory are already exact at the classical level, as it is the case
for the $\cN=2$ prepotential in Calabi-Yau compactifications of M-theory, the following four possibilities can in principle occur,
such that the fields in the consistent truncation do not contribute at one-loop:
The massive modes
\begin{itemize}
 \item are uncharged.
 \item arrange in long multiplets, if the R-symmetry is not gauged.
 \item come in real (non-chiral) representations.
 \item cancel non-trivially between different multiplets.
\end{itemize}
The contributions of long multiplets indeed cancel as one can explicitly check
by using \autoref{t:CS_correct} and \autoref{tab:long_mult} for the Minkowski case. This is related to the fact that they have the structure of
special $\cN=4$ multiplets,
that induce no corrections to the Chern-Simons terms.
For Minkowski space we display the two existing long multiplets in \autoref{tab:long_mult}.
\begin{table}
\begin{center}
\begin{tabular}{lrclr}
\hhline{==~==}
\multicolumn{2}{l}{long gravitino multiplet}&&\multicolumn{2}{l}{long vector multiplet}\TTstrut\BBstrut\\
\hhline{==~==}
field type & $(s_1,s_2)$ && field type & $(s_1,s_2)$ \TTstrut\BBstrut \\
\cline{1-2}\cline{1-2}\cline{4-5}\cline{4-5}
1  gravitino & $(1,\frac{1}{2})$ && 1 vector & $(\frac{1}{2},\frac{1}{2})$ \TTstrut\BBstrut\\
2 tensors & $2 \times (1,0)$ && \multirow{2}{*}{4 fermions} & $2 \times (\frac{1}{2},0)$  \TTstrut\BBstrut\\
2 vectors & $2 \times (\frac{1}{2},\frac{1}{2})$ && & $ 2 \times (0,\frac{1}{2})$ \TTstrut\BBstrut\\
\multirow{2}{*}{5 fermions} & $4 \times (\frac{1}{2},0)$ && 4 scalars & $4 \times (0,0)$ \TTstrut\BBstrut \\
\hhline{~~~==}
& $(0,\frac{1}{2})$ &&& \TTstrut\BBstrut\\
2 scalars & $2 \times (0,0)$ &&& \TTstrut\BBstrut  \\
\hhline{==~~~}
\end{tabular}
\end{center}
\caption{Long multiplets of $\cN=2$ supersymmetry in five-dimensional Minkowski space. The fields are labeled by their spins under $SU(2)\times SU(2)$.}
\label{tab:long_mult}
\end{table}
Also non-chiral multiplets do not contribute since they are parity-invariant
in contrast to the Chern-Simons terms.

After these general considerations let us now turn to some examples. Consider the M-theory reduction on $SU(2)$-structure manifolds of \autoref{su2reduction}.
If the compactification space is also Calabi-Yau, the five-dimensional $\cN=4$ gauged supergravity develops an $\cN=2$ vacuum.
This nicely fits into the general pattern of \autoref{N=4Generalities}. Indeed a Calabi-Yau threefold has $SU(2)$-structure iff its Euler number vanishes.
This can be seen as follows:
A Calabi-Yau threefold has \(SU(3)\) holonomy and thus allows for the existence of one covariantly constant spinor \(\eta^1\).
If the manifold has in addition vanishing Euler number, it follows from the Poincar\'e-Hopf theorem that there exists a nowhere-vanishing vector field \(K^1\).
With this ingredients it is possible to construct a second nowhere vanishing spinor \(\eta^2 = (K^1)^m \gamma_m \eta^1\),
such that the structure group is reduced to $SU(2)$.
This can also be seen without reference to spinors \cite{KashaniPoor:2013en}.
By acting with the complex structure \(J\) on \(K^1\) one obtains a second vector field \(K^2 = J K^1\) and after writing \(J\) and the holomorphic three-form \(\Omega\) as
\begin{equation}
J = J^3 + \tfrac{i}{2} K \wedge \bar K \,,\quad \Omega = K \wedge (J^1 + i J^2) \,,
\end{equation}
it is easy to check that \(K = K^1 + i K^2\) and \(J^a\) fulfill the relations \eqref{eq:su2forms}. 
We could now revert the argument and conclude that a \(SU(2)\) structure manifold with
\begin{equation}\label{eq:CYconditions}
\dd J = \dd \Omega = 0 
\end{equation}
is Calabi-Yau and therefore develops vacua with \(\cN = 2\) supersymmetry.
Using the expansions \eqref{eq:su2formsexpansion} of \(K\) and \(J^a\) we can translate \eqref{eq:CYconditions} into conditions on the five-dimensional fields,
\begin{equation}\begin{aligned}\label{eq:5dCYconditions}
(t^1_I + \tau t^2_I)(\zeta^1_J + i \zeta^2_J) \eta^{IJ} &= 0 \,, \\
(T^K_{1I} + \tau T^K_{2I}) (\zeta^1_J + i \zeta^2_J) \eta^{IJ} &= 0 \,, \\
e^{\rho_4/2} T^J_{iI} \zeta^3_J &= \epsilon_{ij} t^j_I e^{\rho_2} \,.
\end{aligned}\end{equation}
These relations have to be used in the analysis of the spontaneous supersymmetry breaking  to \(\cN = 2\) vacua.
In \autoref{app:contracted_gaugings} we use these conditions in order to derive the contracted embedding tensors \eqref{e:dressed_gaugings}
for Calabi-Yau manifolds with vanishing Euler number. Note that the expressions in \autoref{app:contracted_gaugings} still
suffer from scalar redundancies, and it is hard
to eliminate the latter in general using the Calabi-Yau conditions. However, for the special example of the Enriques Calabi-Yau we were able to do so.
Thus we can derive the full spectrum by inserting the contracted embedding tensors into the results of \autoref{sec:action},
and we will actually do so in the next subsection.
What we will find is that the one-loop Chern-Simons terms do indeed cancel (as in the genuine effective theory),
although very trivially, since there are simply no modes in the theory that are charged under a massless vector.
In fact we think that this might be the generic case for Calabi-Yau manifolds because of the following two
heuristic arguments:
\begin{itemize}
 \item Since a Calabi-Yau manifold has no isometries if the holonomy is strictly $SU(3)$, one would think that the
`KK-vectors' become massive and the massive modes are not charged under massless gauge symmetries.
In particular, the vectors $G^i$ in the ansatz for the metric \eqref{e:su(2)_metric}
\begin{equation}
\dd s^2_{11} = g_{\mu\nu} \dd x^\mu \dd x^\nu + g_{ij} (v^i + G^i)(v^j + G^j) + g_{mn} \dd x^m \dd x^n
\end{equation}
should acquire masses.
\item For Calabi-Yau manifolds with $\chi =0$ and vanishing gaugings $\xi_M$ there are no charged tensors. In fact,
using the Calabi-Yau relations from \eqref{eq:5dCYconditions} it is easy to show that for such manifolds
we have $\xi^{MN}\tensor{\xi}{_N^P}=0$. Applying also the quadratic constraints to \eqref{bos_N=4action} the vanishing of tensor charges is immediate.
Note that the contributions of tensors was a crucial ingredient in \cite{Grimm:2014soa}, where non-vanishing one-loop Chern-Simons terms
appeared in $\cN=4\rightarrow\cN=2$ supergravity breaking to Minkowski vacua.
\end{itemize}
If massive modes carry no charges under massless vectors in general,
our approach via one-loop Chern-Simons terms imposes no restrictions on the consistent truncation to yield also a proper effective theory.

Let us now turn to the second example of partial supergravity breaking in the context of consistent truncations, type IIB supergravity on a
squashed Sasaki-Einstein manifold, which is discussed in \autoref{sec:ex2} in greater detail.
The geometrical reduction to $\cN=4$ gauged supergravity in five dimensions was carried out in \cite{Cassani:2010uw,Liu:2010sa,Gauntlett:2010vu}
and proceeds similarly 
to the M-theory $SU(2)$-structure reduction of \autoref{su2reduction}. Again the theory admits $\cN=2$ vacua, which however now constitute
AdS backgrounds with gauged R-symmetry. Although it is not really clear if the concept of effective field theory makes sense on such backgrounds,
we nevertheless integrate out massive modes. Surprisingly the contributions to the gauge one-loop Chern-Simons term cancel in a  
non-trivial way between
different multiplets. We nevertheless find a non-vanishing correction to the gravitational Chern-Simons term.
It would be nice to find an interpretation for this result.

\subsection{First example: M-theory on the Enriques Calabi-Yau}\label{sec:ex1}

In this subsection we analyze in detail the spectrum of M-theory on the Enriques Calabi-Yau around the $\cN=2$ vacuum of the $\cN=4$
gauged supergravity using the results of \autoref{sec:action}.
The precise expressions for the embedding tensors in the standard form of $\cN =4$ gauged supergravity
and their contractions with the coset representatives for Calabi-Yau manifolds with $SU(2)$-structure are given
in \autoref{app:contracted_gaugings}.
However, as already mentioned, these quantities still contain redundancies from scalar fields, which should be eliminated
by using of the Calabi-Yau conditions \eqref{eq:5dCYconditions} in order to analyze the setup with the tools of \autoref{sec:action}.
Consequently we focus on the special case of the
Enriques Calabi-Yau, where we were able to remove the redundancies.
In the following we derive the spectrum and gauge symmetry in the vacuum of the $SU(2)$-structure reduction
and compare the results to the known Calabi-Yau effective theory.
Besides the fact that the former yields massive states, which are absent in the latter, 
the consistent truncation turns out to lack one vector multiplet and one hypermultiplet at the massless level 
compared to the effective theory of the Enriques, analogous to the results in \cite{KashaniPoor:2013en}.
Taking the missing massless vector into account the classical Chern-Simons terms of both
theories may coincide in principle.
Corrections at one-loop to the Chern-Simons terms vanish trivially, since there are no modes charged under the massless vectors.

The gauged supergravity embedding tensors $f_{MNP}$, $\xi_{MN}$ of M-theory on the Enriques Calabi-Yau
are evaluated by inserting the expressions \eqref{e:torsion_classes} into \eqref{eq:su2embeddingtensors}. In the standard basis, where
$\eta$ takes the form $\eta = (-1,-1,-1,-1,-1,+1,\dots , +1)$, they read
\begin{align}
 &f_{135} = f_{245} = f_{815} = f_{925} = -f_{13\, 10} = -f_{24\, 10} = -f_{81\, 10} = -f_{92\, 10}=\frac{1}{\sqrt 2} \nn \\
 &f_{635} = f_{745} = f_{865} = f_{975} = -f_{63\, 10} = -f_{74\, 10} = -f_{86\, 10} = -f_{97\, 10}= -\frac{1}{\sqrt 2} \nn \\
 &\xi_{13} = \xi_{24} = \xi_{81} = \xi_{92} = 
 -\xi_{63} = -\xi_{74} = -\xi_{86} = -\xi_{97} = \frac{1}{\sqrt 2} \, .
\end{align}
As can be inferred form the covariant derivative \eqref{gen_cov_der}, the gauged $SO(5,n)$ symmetry generators $t_{MN}$ are given by
(modulo normalization of the generators)
\begin{align}
 &t_1 := t_{15} + t_{1\, 10} + t_{65} + t_{6\, 10}\, , & t_2 := t_{25} + t_{2\, 10} + t_{75} + t_{7\, 10} \, , \nn \\
 & t_3 := t_{35} + t_{3\, 10} + t_{85} + t_{8\, 10} \, , 
 &t_4 := t_{45} + t_{4\, 10} + t_{95} + t_{9\, 10} \, , \nn \\
 & t_5 := t_{13} + t_{24} + t_{18} + t_{29} + t_{63} + t_{74} + t_{68} + t_{79} \, .
\end{align}
Since all commutators vanish, as one can check easily, the gauge group in the $\cN =4$ theory is $\big ( U(1)\big )^5$.

Let us now move to the vacuum. The structure of the embedding tensors contracted with the
coset representatives is derived in \autoref{app:contracted_gaugings}. They read
\begin{align}
 &f_{1,6\,\,\, 3,8\,\,\, 5,10} = f_{2,7\,\,\, 4,9\,\,\, 5,10} = \frac{1}{\sqrt 2} \Sigma^3 \lambda_\xi \nn \\
 &\xi_{1,6\,\,\, 3,8} = \xi_{2,7\,\,\, 4,9} = \lambda_\xi \, ,
\end{align}
where for each index position of the tensors there are two options.
For convenience we define
\begin{align}
 \lambda_\xi := \frac{1}{\sqrt 2}\, e^{-\frac{1}{2}(\rho_2 + \rho_4)}\, \I \, \tau \, .
\end{align}
The rotation to $\xi_{\alpha\beta}$, $f_{\alpha\beta\gamma}$ \eqref{e:xi_trafo}, which is the appropriate basis
to split off the propagating degrees of freedom, gives the non-vanishing components
\begin{align}
 & \xi_{12} = \xi_{34} = 2\,\lambda_\xi \, , & f_{125} = f_{345} = f_{12\, 10} = f_{34\, 10} = \sqrt 2 \,\Sigma^3 \,\lambda_\xi \, .
\end{align}

The spectrum is calculated by inserting the contracted embedding tensors into \eqref{e:vac_lagr} and bringing the terms in the Lagrangian into standard form.
The fields together with their masses and charges are listed in \autoref{t:Enriques}.
\begin{table}
\begin{center}
\begin{tabular}{lrrr}
\hline
\hline
multiplet & mass & charge \TTstrut\BBstrut\\
\hline
\rule[-.1cm]{0cm}{.6cm} 1 real graviton multiplet & \multirow{2}{*}{0} & \multirow{2}{*}{\bf0} \TTstrut\BBstrut\\
\rule[-.3cm]{0cm}{.6cm} \((2, 2 \times \tfrac{3}{2}, 1)\) && \TTstrut\BBstrut\\
\hline
\rule[-.1cm]{0cm}{.6cm} 9 real vector multiplets & \multirow{2}{*}{0} & \multirow{2}{*}{\bf0} \TTstrut\BBstrut\\
\rule[-.3cm]{0cm}{.6cm}\((1, 2 \times \tfrac{1}{2}, 0)\) && \TTstrut\BBstrut\\
\hline
\rule[-.1cm]{0cm}{.6cm}  11 real hypermultiplets & \multirow{2}{*}{0} & \multirow{2}{*}{\bf0} \TTstrut\BBstrut\\
\rule[-.3cm]{0cm}{.6cm} \((2 \times \tfrac{1}{2}, 4 \times 0)\) && \TTstrut\BBstrut\\
\hline
\rule[-.1cm]{0cm}{.6cm} 1 complex gravitino multiplet & \multirow{2}{*}{\(m\)} & \multirow{2}{*}{\bf0} \TTstrut\BBstrut\\
\rule[-.3cm]{0cm}{.6cm} \(\left((1,\tfrac{1}{2}), 2 \times (1, 0), 2 \times (\tfrac{1}{2}, \tfrac{1}{2}), 4 \times (\tfrac{1}{2}, 0), (0, \tfrac{1}{2}), 2 \times (0,0)\right)\) && \TTstrut\BBstrut\\
\hline
\rule[-.1cm]{0cm}{.6cm} 1 real vector multiplet & \multirow{2}{*}{\(2mc\)} & \multirow{2}{*}{\bf0} \TTstrut\BBstrut\\
\rule[-.3cm]{0cm}{.6cm} \(\left((\tfrac{1}{2}, \tfrac{1}{2}), 2 \times (\tfrac{1}{2},0)\right)\) &&\TTstrut\BBstrut \\
\hline
\rule[-.1cm]{0cm}{.6cm}  1 complex hypermultiplet & \multirow{2}{*}{\(2m\)} & \multirow{2}{*}{\bf0} \TTstrut\BBstrut\\
\rule[-.3cm]{0cm}{.6cm}  \(\left((\tfrac{1}{2}, 0), 2 \times (0,0)\right)\) && \TTstrut\BBstrut\\
\hline
\hline
\end{tabular}
\end{center}
\caption{Spectrum of the $SU(2)$-structure reduction of M-theory on the Enriques Calabi-Yau.}
\label{t:Enriques}
\end{table}
The modes are classified according to their mass, charges under the massless vectors and their
representations under $SO(3)\cong SU(2)$ or $SO(4)\cong SU(2) \times SU(2)$, the massless and massive little groups, respectively.
Fermions in complex multiplets are Dirac, while fermions in real multiplets are taken to be symplectic Majorana.
We set $m=\sqrt 2 \, \Sigma^{2}\lambda_\xi$
and $c=\frac{(1+\Sigma^{-6})^{3/2}}{(1+\Sigma^{-12})^{1/2}}$ .

The massless multiplets are uncharged and consistent with the proper Calabi-Yau effective theory apart from one missing vector multiplet and
one hypermultiplet.
More precisely, the Enriques Calabi-Yau has Hodge numbers $h^{1,1}= h^{2,1} = 11$.
In the effective action of M-theory on Calabi-Yau threefolds one finds $h^{1,1}-1$ vector multiplets and
$h^{2,1}+1$ hypermultiplets, while for our consistent truncation on the Enriques Calabi-Yau we find only 9 vector multiplets and 11
hypermultiplets.
This resembles the results in \cite{KashaniPoor:2013en} where the same field content was missing for the analog type IIA setup.
Geometrically the corresponding missing harmonic forms are captured by
$SU(2)$-doublets, which we discarded in the reduction of \autoref{su2reduction}.
As explained, the doublets correspond to $\cN = 4$ gravitino multiplets, for which
no coupling to standard $\cN =4$ gauged supergravity is known.
Having discussed the massless modes in the vacuum, we turn to the massive spectrum.
We find one long gravitino multiplet, one vector multiplet and one hypermultiplet.
Interestingly no massive field is charged under a massless $U(1)$ gauge symmetry.
For the massive tensors this has already been established on general grounds in the last subsection.
Thus we conclude that for the Enriques Calabi-Yau the Chern-Simons terms \eqref{e:standard_CS} are trivially not corrected by loops
of fermions or tensors, since there are no charged modes in the truncation.

Finally let us also comment on the classical Chern-Simons terms in the reduction. We denote the ten massless vectors in the vacuum of the consistent truncation
by $\tilde A^1_\mu$, $\tilde A^2_\mu$, $\tilde A^\mathfrak{a}_\mu$ with $\mathfrak{a} =1,\dots,8$. The $\tilde A^\mathfrak{a}_\mu$ originate from the
$E_8$ nature of the Enriques surface. The classical Chern-Simons coefficients are found to be
\begin{align}
& k^{\textrm{trunc}}_{121}=2\sqrt 2 \, , & k^{\textrm{trunc}}_{1\mathfrak{a}\mathfrak{a}}=2 \, ,
\end{align}
all others vanish. In the familiar Calabi-Yau effective action the Chern-Simons coefficients reproduce the intersection numbers
of the manifold. For the Enriques Calabi-Yau they read in a suitable basis
\begin{align}\label{e:intersec_enriques}
 & k^{\textrm{eff}}_{123}=1 \, , & k^{\textrm{eff}}_{1\mathfrak{a}\mathfrak{b}}=A^{E_8}_{\mathfrak{a}\mathfrak{b}} \, ,
\end{align}
where $A^{E_8}$ denotes the Cartan matrix of $E_8$.
If we assume that the missing vector $\tilde A^{3}$ appears together with $\tilde A^{1}$ and $\tilde A^{2}$ in a Chern-Simons term with coefficient
\begin{align}
  k^{\textrm{miss}}_{123} \neq 0 \, ,
 \end{align}
 we can define
 \begin{align}
  &\hat A^1_\mu := \tilde A^1_\mu \ , &&
  \hat A^1_\mu := \tilde A^1_\mu \, , &&
  \hat A^3_\mu := \sqrt 2\, \tilde A^1_\mu + k^{\textrm{miss}}_{123} \cdot \tilde A^3_\mu \, , 
 \end{align}
 such that in this basis we obtain Chern-Simons coefficients 
 \begin{align}
  & k_{123} =1 \, ,  & k_{1\mathfrak{a}\mathfrak{a}}=2 \, .
 \end{align}
The first one matches with \eqref{e:intersec_enriques}. Concerning the second term we note that
the Cholesky decomposition of $A^{E_8}$ ensures that there exists a field redefinition for the $\hat A^\mathfrak{a}_\mu$
represented by a matrix $T$, which fulfills
\begin{align}
 T^T T = \frac{1}{2} A^{E_8} \, .
\end{align}
It is easy to check that under this redefinition
$k_{1\mathfrak{a}\mathfrak{a}}$ goes to $k^{\textrm{eff}}_{1\mathfrak{a}\mathfrak{b}}$.
These considerations can also be interpreted as a proposition for the Chern-Simons coefficient, 
which involve the missing massless vector $\tilde A^{3}$,
namely $k^{\textrm{miss}}_{123} \neq 0$. It should be reproduced by the $SU(2)$-doublets.
 
We conclude that for the Enriques Calabi-Yau,
apart from the missing vector multiplet and hypermultiplet, the effective theory of the consistent truncation is consistent
with the genuine Calabi-Yau effective action, since it is in principle possible to match the classical Chern-Simons terms of both sides, and more importantly
corrections at one-loop are absent in the consistent truncation, since massive modes do not carry any charges.
As we think that this is the case for generic Calabi-Yau manifolds with vanishing Euler number, 
the analysis of the Chern-Simons terms reveals no restrictions for the consistent truncation to also
yield a proper effective action. We believe that this conclusion changes significantly, if the 
internal space has isometries and there are massive modes charged under 
massless vectors. We will turn to an example that has these features next. 

\subsection{Second example: Type IIB supergravity on a squashed Sasaki-Einstein manifold}\label{sec:ex2}

In the following we study a second example of partial supergravity breaking in the context 
of consistent truncations that features a massive spectrum charged under a massless vector.
More precisely, we consider type IIB supergravity
on a squashed Sasaki-Einstein manifold with 5-form flux. This setup admits a consistent truncation to $\cN = 4$ gauged supergravity in five dimensions,
which has two vacua, one which breaks supersymmetry completely and one which is $\cN =2$ AdS. We focus on the latter.
Since the theory in the broken phase can be described with the results of \autoref{sec:action},
we proceed along the lines of the last subsection and derive the spectrum and Chern-Simons terms.
The field content turns out to be consistent with \cite{Cassani:2010uw,Liu:2010sa,Gauntlett:2010vu}.
Although there are massive modes charged under the gauged R-symmetry in the vacuum, their corrections to the \emph{gauge} Chern-Simons term at one-loop
cancel exactly. However, the \emph{gravitational} one-loop Chern-Simons term does not vanish.

In \cite{Cassani:2010uw} it was shown that in a consistent truncation of type IIB supergravity on a 
squashed Sasaki-Einstein manifold to 5D $\cN=4$ gauged supergravity
the non-vanishing embedding tensors $f_{MNP}$, $\xi_{MN}$ take the form
\begin{align}
 & f_{125}=f_{256}=f_{567}=-f_{157}=-2 \, , \nn \\
& \xi_{12}=\xi_{17}=-\xi_{26}=\xi_{67}=-\sqrt 2 k \, , &\xi_{34}= -3\sqrt 2 \, , 
\end{align}
where $k$ denotes 5-form flux on the internal manifold.
They encode the gauging of the group $\textrm{Heis}_3 \times U(1)_R$, where a $U(1)_R$ is a subgroup of the R-symmetry group.
The theory admits a vacuum that preserves $\cN =2$ supersymmetry.
If we for simplicity fix the RR-flux to $k=2$, we can use the expressions for the scalar VEVs in \cite{Cassani:2010uw}
to derive the contracted embedding tensors \eqref{e:dressed_gaugings}
\begin{align}
& f_{125}=f_{675}=-f_{175}=-f_{625}=2 \, , \nn \\
& \xi_{12}=\xi_{67}=-\xi_{17}=-\xi_{62}=-2 \sqrt 2  \, , &\xi_{34}= -3\sqrt 2 \, . 
\end{align}
We can now rotate into the basis of \eqref{e:xi_trafo}, in fact we transform $\xi_{\hat\alpha\hat\beta}$ already into block-diagonal form.
The non-vanishing gaugings $\xi_{\alpha\beta}$, $f_{\alpha\beta\gamma}$ read
\begin{align}
 & \xi_{12} = 4 \sqrt 2 \, , && \xi_{34} = 3 \sqrt 2 \, ,
 & f_{125} = -4
\end{align}
and therefore
\begin{align}
 & \hat \alpha = 1,2,3,4 \, , & \tilde \alpha = 5,6,7 \, .
\end{align}

Carrying out the calculations we find the cosmological constant $\Lambda = -6$ corresponding to an AdS$_5$ background.
Furthermore half of the supersymmetries are broken and the gauge group is reduced
\begin{align}
 \textrm{Heis}_3 \times U(1)_R \rightarrow U(1)_R \, ,
\end{align}
where now the full $U(1)$ R-symmetry of AdS$_5$ is gauged with gauge coupling $g= \sqrt \frac{3}{2}$ .
The full spectrum of the consistent truncation in the vacuum is depicted in \autoref{t:SE},
where we consulted the categorization of \cite{Ceresole:1999zs}.
The fields are classified
according to their mass, charge under $U(1)_R$ with coupling $g$ and their
representation under the $SU(2)\times SU(2)$ part of the maximal compact subgroup of $SU(2,2|1)$.
\begin{table}
\begin{center}
\begin{tabular}{llrr}
\hline\hline
multiplet & representation & mass & charge   \\
\hline
\multirow{4}{*}{1 real graviton multiplet}
& $(1,1)$ & 0 & 0\TTstrut\BBstrut\\
& $(1,\frac{1}{2})$ & 0 & -1\TTstrut\BBstrut\\
& $(\frac{1}{2},1)$ & 0 & +1\TTstrut\BBstrut\\
& $(\frac{1}{2},\frac{1}{2})$ & 0 & 0\TTstrut\BBstrut\\
\hline
\multirow{3}{*}{1 complex hypermultiplet}
& $(\frac{1}{2},0)$ & 3/2 & +1\TTstrut\BBstrut\\
& $(0,0)$ & -3 & +2\TTstrut\BBstrut\\
& $(0,0)$ &  0& 0\TTstrut\BBstrut\\
\hline
\multirow{6}{*}{1 complex gravitino multiplet}
& $(\frac{1}{2},1)$ & -5 & +1\TTstrut\BBstrut\\
& $(\frac{1}{2},\frac{1}{2})$ & 8 & 0\TTstrut\BBstrut\\
& $(0,1)$ & 3 & +2\TTstrut\BBstrut\\
& $(0,1)$ & 4 & 0\TTstrut\BBstrut\\
& $(0,\frac{1}{2})$ & -5/2 & +1\TTstrut\BBstrut\\
& $(0,\frac{1}{2})$ & -7/2 & +3\TTstrut\BBstrut\\
\hline
\multirow{9}{*}{1 real vector multiplet}
& $(\frac{1}{2},\frac{1}{2})$ & 24 & 0\TTstrut\BBstrut\\
& $(\frac{1}{2},0)$ & 9/2 & -1\TTstrut\BBstrut\\
& $(0,\frac{1}{2})$ & 9/2 & +1\TTstrut\BBstrut\\
& $(0,\frac{1}{2})$ & 11/2 & -1\TTstrut\BBstrut\\
& $(\frac{1}{2},0)$ & 11/2 & +1\TTstrut\BBstrut\\
& $(0,0)$ & 12 & 0\TTstrut\BBstrut\\
& $(0,0)$ & 21 & -2\TTstrut\BBstrut\\
& $(0,0)$ & 21 & +2\TTstrut\BBstrut\\
& $(0,0)$ & 32 & 0\TTstrut\BBstrut\\
\hline\hline
\end{tabular}
\end{center}
\caption{Spectrum of type IIB supergravity on a squashed Sasaki-Einstein manifold in the $\cN=2$ vacuum corresponding to an AdS$_5$ background.}
\label{t:SE}
\end{table}

For our example we find at the classical level\footnote{We do not account for the classical gravitational Chern-Simons term.} 
\begin{align}
 k_{000}^{\textrm{class}}= 4\, \sqrt{\frac{2}{3}} \, .
\end{align}
In order to calculate the quantum corrections, we again use \autoref{t:CS_correct} with the understanding that representations of
$SU(2)\times SU(2) \subset SU(2,2|1)$ in AdS
contribute in the same way as representations of $SU(2)\times SU(2) \cong SO(4)$ in the Minkowski case.
Although the results of \autoref{t:CS_correct}, derived in \cite{Bonetti:2013ela}, were originally calculated in a Minkowski background, we believe
that they are applicable to AdS as well, since they can be derived solely from anomalies.
Remarkably, the one-loop corrections of the massive charged modes to the gauge Chern-Simons term cancel in a highly non-trivial way,
while the gravitational Chern-Simons term does receive corrections
\begin{align}
 & k_{000}^{\textrm{1-loop}}=0 \, , & k_{0}^{\textrm{1-loop}}=72 \,\sqrt{\frac{3}{2}} \, .
\end{align}
Note that the index zero is now meant to refer to the remaining massless $U(1)_R$ in the vacuum rather than to $A^0$ in the $\cN = 4$ theory.

The interpretation of these results is not as clear as in the last subsection concerning the Enriques Calabi-Yau. Indeed, 
the naive notion of an effective field theory on AdS backgrounds will not be well-defined, if the AdS radius is linked to 
the size of the internal space. We nevertheless think that the non-trivial vanishing of the gauge
one-loop Chern-Simons term is not accidental and should have a clear interpretation. Related to that, it would also be interesting
to find connections to other consistent truncations. The simplest example is certainly the $\cN=8$ consistent truncation to massless modes of
type IIB supergravity on the five-sphere \cite{Cvetic:2000nc}, which is a special Sasaki-Einstein manifold.

\section{Conclusions}
In this paper we studied spontaneous breaking of five-dimensional $\cN=4$ gauged supergravity. We analyzed the theory in the broken phase around
the vacuum by deriving the spectrum including charges and masses, as well as Chern-Simons terms and the cosmological constant.
Special focus was put on setups with $\cN=2$ vacua, since examples of this type arise in consistent truncations
of string theory and supergravity.

As consistent truncations of non-Calabi-Yau reductions are exploited for phenomenological investigations, it is a crucial
task to provide necessary conditions for them to yield valid effective actions upon integrating out massive modes. 
Consequently, we required the one-loop Chern-Simons terms in consistent truncations, induced by massive charged modes,
to match their counterparts of the genuine effective action.

As a first example, we considered consistent truncations of M-theory on $SU(2)$-structure manifolds to $\cN=4$
gauged supergravity in five dimensions. The geometrical ansatz includes
$SU(2)$-singlets and triplets, while doublets are excluded, since they lead to gravitino multiplets, for which no consistent coupling
to supergravity is known. We first derived the general five-dimensional $\cN=4$ action for this ansatz including a non-trivial 
flux background. The vacua of a specific class of $SU(2)$-structure manifolds, namely Calabi-Yau manifolds with vanishing Euler number, were then analyzed in in greater detail.
They constitute $\cN =2$ vacua and can therefore be analyzed with the supergravity breaking mechanism described in the first part
of the paper. It turned out that one can generally miss at the massless level vector multiplets and hypermultiplets, 
which are captured by the $SU(2)$-doublets that we omitted in our ansatz.
At the quantum level the requirement for having a proper effective theory necessitates the vanishing of one-loop corrections 
to the Chern-Simons terms, since the relevant parts of the genuine effective action are classically exact.
Indeed, by analyzing Enriques Calabi-Yau manifold as an example, we found that no massive charged modes 
appear and the one-loop Chern-Simons terms cancel trivially.
We argued that this might be the case for general consistent truncations on Calabi-Yau spaces.
Accordingly, we claim that, apart from the missing massless degrees of 
freedom, the Chern-Simons terms provide no immediate contradiction to 
deriving proper effective theories from consistent truncations in the Calabi-Yau case. 
This might be traced back to the fact that a Calabi-Yau manifold with $SU(3)$ holonomy
has no continuous isometries and therefore no massless Kaluza-Klein vector. 

It is interesting to speculate on $SU(2)$-structure reductions with isometries. 
In such situations one finds massless Kaluza-Klein vectors gauging 
massive modes. Integrating out the massive fields one expects to 
find one-loop Chern-Simons terms as known from circle reductions \cite{Intriligator:1997pq,Bonetti:2012fn,Bonetti:2013ela}.
Here it appears to be crucial to distinguish the case of integrating 
out infinitely many modes from the case of considering only a finite truncation.
In other words, we suspect that in this case the consistent truncation 
might not yield an effective theory that matches the genuine effective action 
of the complete reduction. It would be interesting to find non-trivial examples 
for this situation. 

We also investigated a second example of partial supergravity breaking accompanying consistent truncations
where the internal manifold has isometries. More precisely, we considered
type IIB supergravity on a squashed Sasaki-Einstein manifold with RR-flux. This example can be interpreted 
also as a special case of a warped $SU(2)$-structure reduction of
M-theory \cite{Gauntlett:2004zh,Gauntlett:2006ai,Gauntlett:2007sm,OColgain:2011ng,Sfetsos:2014tza}.
The squashed Sasaki-Einstein reductions give also an $\cN=4$ gauged supergravity in five dimensions, which
now admits an $\cN=2$ AdS vacuum.
This time there indeed appear massive states charged under the gauged $U(1)$ R-symmetry.
For the gravitational Chern-Simons term we found non-vanishing one-loop contributions, however, remarkably the corrections to the gauge Chern-Simons term
cancel in a non-trivial way. While one might question the existence of a proper effective theory for these 
AdS backgrounds, the cancelations in the Chern-Simons terms are intriguing
It would be nice to gain a deeper understanding of this fact.

Let us close our conclusions by pointing out that an analysis similar to 
the one of this paper can be carried out for M-theory reductions on 
real eight-dimensional manifolds. In this case the effective theory 
will be three-dimensional, but can also contain one-loop Chern-Simons 
terms that see the massive spectrum. It would be interesting, for example, 
to consider M-theory on Spin(7) manifolds or Calabi-Yau fourfolds with 
vanishing Euler number. In these cases one finds additionally an enlarged 
structure group and therefore a three-dimensional gauged supergravity 
theory with partially supersymmetry breaking vacua.

\subsubsection*{Acknowledgments}

We would like to thank Federico Bonetti, Eoin Colg\'ain, Jan Louis, Tom Pugh, and Hagen Triendl for interesting discussions and comments.
This work was supported by a research grant of the Max Planck Society.

\appendix

\section{Conventions and identities} \label{App-Conventions}

We shortly state the conventions of differential geometry used in this paper.
Curved five-dimensional spacetime indices are denoted by Greek letters $\mu,\nu,\dots$. Antisymmetrizations
of any kind are always done with weight one, i.e.~include a factor of $1/n!\,$.
We use the $(-,+,+,+,+)$ convention for the five-dimensional metric $g_{\mu\nu}$, and we adopt the negative sign in front of the Einstein-Hilbert term.
Moreover we set
\begin{align}
 \kappa^2  = 1 \, .
\end{align}
The Levi-Civita tensor with curved indices $\epsilon_{\mu\nu\rho\lambda\sigma}$ reads
\begin{align}
 \epsilon_{01234} = + e \, , \qquad  \, , \epsilon^{01234} = - e^{-1}\, ,
\end{align}
where $e = \sqrt{-\det g_{\mu\nu} }\,$.

The five-dimensional spacetime gamma matrices are denoted by $\gamma_\mu$ and satisfy
\begin{align}
 \lbrace \gamma_\mu , \gamma_\nu \rbrace = 2 g_{\mu\nu} \, .
\end{align}
Antisymmetrized products of gamma matrices are defined as
\begin{align}
 \gamma_{\mu_1 , \dots , \mu_k } := \gamma_{[ \mu_1}\gamma_{\mu_2}\dots \gamma_{\mu_k ]}\, . 
\end{align}
The convention for the charge conjugation matrix $C$ is such that
\begin{align}
 C^T = -C = C^{-1}
\end{align}
and it fulfills
\begin{align}
 C \gamma_\mu C^{-1} = ( \gamma_\mu )^T \, .
\end{align}
All massless spinors in this paper are meant to be symplectic Majorana, that is in the $\cN=4$ theory they are subject to the condition
\begin{align}
 \bar \chi^i := (\chi_i)^\dagger \gamma_0 = \Omega^{ij} \chi_j^T C \, ,
\end{align}
where $i,j =1, \dots , 4$ and $\Omega^{ij}$ is the symplectic form of $USp(4)$ defined in \eqref{e:properties_omega}.
In the $\cN=2$ theory the symplectic Majorana condition reads
\begin{align}
 \bar \chi^\alpha := (\chi_\alpha)^\dagger \gamma_0 = \varepsilon^{\alpha\beta} \chi_\beta^T C \, ,
\end{align}
where $\alpha,\beta = 1,2$, $\varepsilon^{\alpha\beta}$ is the two-dimensional epsilon tensor.

The standard Lagrangians of massive fields in five dimensions are useful for explicit manipulations of the results in \autoref{sec:action}.
For a massive tensor charged under a $U(1)$ gauge symmetry we have
\begin{align} \label{standard_complex_tensor}
  e^{-1}\cL_B &= -\frac{1}{4}i c_\BB\, \epsilon^{\mu\nu\rho\sigma\tau}\bar \BB_{\mu\nu}\cD_\rho \BB_{\sigma\tau} 
 -\frac{1}{2}m_\BB \, \bar \BB_{\mu\nu} \BB^{\mu\nu}\, , 
\end{align}
with $\cD_\rho \BB_{\sigma\tau} =   \partial_\rho \BB_{\sigma\tau} - i q_\BB\, A_\rho\, \BB_{\sigma\tau}$.
The quantity $m_\BB>0$ is the physical mass of the complex 
tensor  and $q_\BB$ encodes its charge under $A_\mu$.
Furthermore the representation under the little group is encoded in
\begin{align} \label{cBB_choice}
 c_{\BB} &= + 1 \Leftrightarrow (1,0 )\textrm{ of } SU(2)\times SU(2) \\
 c_{\BB} &= - 1 \Leftrightarrow ( 0,1 ) \textrm{ of } SU(2)\times SU(2)  \, . \nn
\end{align}
A massive, charged Dirac spin-3/2 fermion is described by
\begin{align}
 \label{e:lagrangian_gravitino}
 e^{-1} \cL_\Bpsi = - \bar \Bpsi_\mu \gamma^{\mu\nu\rho}  \cD_\nu \Bpsi_{\rho} + 
  c_{\Bpsi}\,  m_\Bpsi \, 
 \bar \Bpsi_\mu \gamma^{\mu\nu} \Bpsi_{\nu }
 \, , 
\end{align}
where $\cD_\nu \Bpsi_{\rho} = \partial_\nu \Bpsi_{\rho} - i q_\Bpsi A_{\nu} \Bpsi_\rho$ and  
\begin{align} \label{cBpsi_choice}
 c_{\Bpsi} &= + 1 \Leftrightarrow (\tfrac{1}{2}, 1 ) \textrm{ of } SU(2)\times SU(2)\, ,  \\
 c_{\Bpsi} &= - 1 \Leftrightarrow (1 , \tfrac{1}{2}) \textrm{ of } SU(2)\times SU(2)  \, . \nn
\end{align}
Finally the Lagrangian of a massive, charged Dirac spin-1/2 fermion reads
\begin{align}
\label{e:lagrangian_1/2}
 e^{-1} \cL_\Blambda = -\bar \Blambda \slashed \cD \Blambda +  c_{\Blambda}  m_\Blambda \bar \Blambda \Blambda \, ,
\end{align}
with $\cD_\nu \Blambda = \partial_\nu \Blambda - i q_\Blambda A_{\nu} \Blambda$ and
\begin{align}  \label{cBlambda_choice}
 c_{\Blambda} &= + 1 \Leftrightarrow (\frac{1}{2}, 0 )\textrm{ of } SU(2) \times SU(2) \\
 c_{\Blambda} &= - 1 \Leftrightarrow (0 , \frac{1}{2})\textrm{ of } SU(2) \times SU(2) \, . \nn
\end{align}

\section{The coset representative and contracted embedding tensors for \texorpdfstring{$SU(2)$}{SU(2)}-structure manifolds}\label{app:cosetrepr}
\subsection{The coset representative \texorpdfstring{$\cV$}{V}}
From \eqref{eq:scalarmetric} we can extract representatives \(\cV = ({\cV_M}^m, {\cV_M}^a)\) of the coset space
\begin{equation}
\frac{SO(5,n-1)}{SO(5) \times SO(n-1)} \,,
\end{equation}
where \(m = 1,\dots,5\) and \(a = 6,\dots,5+n\) denote \(SO(5)\) and \(SO(n)\) indices.
These coset representatives are related to the scalar metric via
\begin{equation}\label{eq:cosetm}
(M_{MN}) = \cV\cV^T = \cV^m\cV^m + \cV^a\cV^a
\end{equation}
and have to fulfill
\begin{equation}\label{eq:coseteta}
(\eta_{MN}) = -\cV^m\cV^m + \cV^a\cV^a \,.
\end{equation}
Before we can determine \(\cV\), it is necessary to diagonalize \(g_{ij}\) and \(H_{IJ}\). First we observe that \(g_{ij}\) can be expressed as
\begin{equation}
g_{ij} = e^{-\rho_2}k_i^k k_j^l \delta_{kl} \,,
\end{equation}
where \(k = e^{\rho_2/2} (\I\,\tau)^{-1/2}(1,\tau)\).

In \eqref{eq:su2ansatzhodge} we have introduced \(H_{IJ}\) via \(\ast \omega^I = -{H^I}_J \omega^J \wedge \mathrm{vol}_2^{(0)}\) and as described in \cite{Louis:2009dq} it depends only on \(\zeta^a_I\),
\begin{equation}\label{eq:zetaH}
H_{IJ} = 2 \zeta^a_I \zeta^a_J + \eta_{IJ} \,.
\end{equation}
From \eqref{eq:su2forms} and \eqref{eq:su2formsexpansion} one sees that
\begin{equation}
\zeta^a_I \eta^{IJ} \zeta^b_J = -\delta^{ab} \,.
\end{equation}
Therefore \(\zeta^a_I {H^I}_J = -\zeta^a_J\), that means that the three \(\zeta^a_I\) are eigenvectors of \({H^I}_J\) with eigenvalue \(-1\). 
If we now introduce an orthonormal basis \(\xi^\alpha_I\) (\(\alpha = 1,\dots,n-3\)) of the subspace orthogonal to all \(\zeta^a_I\) (i.e.~\(\xi^\alpha_I \eta^{IJ} \xi^\beta_J = \delta^{\alpha\beta}\), \(\zeta^a_I \eta^{IJ} \xi^\beta_J = 0\)), we can write 
\begin{equation}
H_{IJ} = \zeta^a_I \zeta^a_J + \xi^\alpha_I \xi^\alpha_J,
\end{equation}
since we deduce from \eqref{eq:zetaH} that the \(\xi^\alpha_I\) are eigenvectors of \({H^I}_J\) with eigenvalue \(+1\).
Moreover it follows that \(\xi^\alpha_I \xi^\alpha_J = \zeta^a_I \zeta^a_J + \eta_{IJ}\) and so
\begin{equation}
\eta_{IJ} = - \zeta^a_I \zeta^a_J + \xi^\alpha_I \xi^\alpha_J \,.
\end{equation} 
We can shorten the notation by defining
\begin{equation}\label{eq:combinedzetaxi}
E^\cI_I = (\zeta^a_I, \xi^\alpha_J) \,, \quad \cI = (a, \alpha) \,,
\end{equation}
which allows us to write
\begin{equation}
H_{IJ} = E^\cI_I E^\cJ_J \delta_{\cI\cJ}
\quad\mathrm{and}\quad
\eta_{IJ} = E^\cI_I E^\cJ_J \eta_{\cI\cJ} \,,
\end{equation}
with \(\eta_{\cI\cJ} = \mathrm{diag}(-1,-1,-1;+1,\dots,+1)\).

After this preparation we are able to write down \(\cV\),
\begin{equation}\begin{aligned}\label{eq:su2cosetrepr}
{\cV_i}^j &= e^{\rho_4/2} k_i^j \,, \\
{\cV_i}^\bj &= e^{-\rho_4/2} \delta^{j\bj} (k^{-1})_j^k(\epsilon_{ki} \gamma + \tfrac{1}{2}c_{kI}c_i^I) \,, \\
{\cV_i}^\cI &= -{E_I}^\cI c_i^I \,, \\
{\cV_\bi}^\bj &= e^{-\rho_4/2}\delta^{j\jb} \delta_{i\ib} (k^{-1})_j^i \,, \\
{\cV_I}^\bi &= e^{-\rho_4/2}\delta^{i\ib} (k^{-1})_i^j c_{jI} \,, \\
{\cV_I}^\cI &= {E_I}^\cI \,,
\end{aligned}\end{equation}
such that
\begin{equation}\label{eq:cosetnondiagm}
M_{MN} = (\cV \cV^T)_{MN} = {\cV_M}^i {\cV_N}^i + {\cV_M}^\ib {\cV_N}^\ib + {\cV_M}^\cI {\cV_N}^\cI \,,
\end{equation}
and
\begin{equation}\label{eq:cosetnondiageta}
\eta_{MN} = 2 \delta_{i\ib} {\cV_M}^i {\cV_N}^{\ib} + \eta_{\cI\cJ}{\cV_M}^\cI {\cV_N}^\cJ \,.
\end{equation}

In the end it is necessary to split \eqref{eq:su2cosetrepr} into \(\cV^m\) and \(\cV^a\), which corresponds to bringing \eqref{eq:cosetnondiageta} in diagonal form.
The result reads
\begin{equation}
{\cV_M}^m =
\begin{pmatrix}
\tfrac{1}{\sqrt 2}\left(-{\cV_M}^1 + {\cV_M}^{\bar1}\right) \\
\tfrac{1}{\sqrt 2}\left(-{\cV_M}^2 + {\cV_M}^{\bar2}\right) \\
{\cV_M}^{\cI = 1,2,3}
\end{pmatrix} \,, \quad
{\cV_M}^a =
\begin{pmatrix}
\tfrac{1}{\sqrt 2}\left({\cV_M}^1 + {\cV_M}^{\bar1}\right) \\
\tfrac{1}{\sqrt 2}\left({\cV_M}^2 + {\cV_M}^{\bar2}\right) \\
{\cV_M}^{\cI \neq 1,2,3}
\end{pmatrix}\,.
\end{equation}
With \eqref{eq:cosetnondiagm} and \eqref{eq:cosetnondiageta} one can easily check that these combinations fulfill \eqref{eq:cosetm} and \eqref{eq:coseteta}.

\subsection{The contracted embedding tensors for Calabi-Yau manifolds with \texorpdfstring{$\chi = 0$}{chi=0}}\label{app:contracted_gaugings}
Using the results from \eqref{eq:su2cosetrepr} we can compute the contractions of the embedding tensors \eqref{eq:su2embeddingtensors} with the coset representatives, as introduced in \eqref{e:dressed_gaugings}.
Hereby we restrict to the special case of Calabi-Yau manifolds with vanishing Euler number and use the relevant relations from \eqref{eq:5dCYconditions} that follow to simplify the resulting expressions.
We also restrict to the case without four-form flux and set \(n = n_I = 0\).

For \(\xi^{mn}\) we find that it takes the general form
\begin{equation}
\xi^{mn} = 
\left( \begin{array}{@{}cc|ccc@{}}
\multicolumn{2}{c|}{\multirow{2}*{${\bf0}_{2\times2}$}} & - & \xi^{1n} & - \\
&& - & \xi^{2n} & - \\
\hline
| & | & \multicolumn{3}{c}{\multirow{3}*{${\bf0}_{3\times3}$}} \\
\xi^{m1} & \xi^{m2} && \\
| & | &&& \\
\end{array}\right) \,,
\end{equation}
where its non-vanishing components are given by
\begin{equation}\begin{aligned}\label{eq:xi12m}
\xi^{1,m=3,4,5} = -\xi^{m1} &= \frac{1}{\sqrt2}e^{-\frac{1}{2}(\rho_2+\rho_4)}\sqrt{\I\,\tau} \,t^{2I} \zeta^{a=1,2,3}_I \,,\\
\xi^{2,m=3,4,5} = -\xi^{m2} &= - \frac{1}{\sqrt2}e^{-\frac{1}{2}(\rho_2+\rho_4)}\frac{1}{\sqrt{\I\,\tau}} \left(t^{1I} + \R\,\tau \,t^{2I}\right) \zeta^{a=1,2,3}_I \,.
\end{aligned}\end{equation}
Similarly we have
\begin{equation}
\xi^{ab} = 
\left( \begin{array}{@{}cc|ccc@{}}
\multicolumn{2}{c|}{\multirow{2}*{${\bf0}_{2\times2}$}} & - & \xi^{6b} & - \\
&& - & \xi^{7b} & - \\
\hline
| & | & \multicolumn{3}{c}{\multirow{3}*{${\bf0}_{(n-2)\times(n-2)}$}} \\
\xi^{a6} & \xi^{a7} && \\
| & | &&& \\
\end{array}\right),
\end{equation}
with\footnote{Note that the indices $n$, defined around \eqref{e:content_vector}, and $\tilde n$, defined around \eqref{eq:su2ansatzintegral},
are related by $n=\tilde n -1$.}
\begin{equation}\begin{aligned}\label{eq:xi12a}
\xi^{6,a=8,\dots, 5+n} = -\xi^{a6} &= \frac{1}{\sqrt2}e^{-\frac{1}{2}(\rho_2+\rho_4)}\sqrt{\I\,\tau} \,t^{2I} \xi^{\alpha=1,\dots,\tilde n-3}_I \,,\\
\xi^{7,a=8,\dots, 5+n} = -\xi^{a7} &= - \frac{1}{\sqrt2}e^{-\frac{1}{2}(\rho_2+\rho_4)}\frac{1}{\sqrt{\I\,\tau}} \left(t^{1I} + \R\,\tau \,t^{2I}\right) \xi^{\alpha=1,\dots,\tilde n-3}_I \,,
\end{aligned}\end{equation}
and finally for the mixed-index part
\begin{equation}
\xi^{ma} = 
\left( \begin{array}{@{}cc|ccc@{}}
\multicolumn{2}{c|}{\multirow{2}*{${\bf0}_{2\times2}$}} & - & \xi^{1a} & - \\
&& - & \xi^{2a} & - \\
\hline
| & | & \multicolumn{3}{c}{\multirow{3}*{${\bf0}_{3\times( n-2)}$}} \\
\xi^{m6} & \xi^{m7} && \\
| & | &&& \\
\end{array}\right) \,,
\end{equation}
where its entries are again given by \eqref{eq:xi12m} and \eqref{eq:xi12a}.

Following the notation introduced in \eqref{eq:combinedzetaxi} we obtain for the non-vanishing components of the contracted \(f_{MNP}\)
\begin{equation}\begin{aligned}
f^{m=1,\cI\cJ} = f^{a=6,\cI\cJ} =& - \frac{1}{\sqrt2}e^{-\frac{1}{2}(\rho_2+\rho_4)}\sqrt{\I\,\tau} \Bigl(T^J_{1K} \eta^{KI} + \tfrac{1}{2} t^2 \eta^{IJ} \Bigr) E^\cI_I E^\cJ_J \,, \\
f^{m=2,\cI\cJ} = f^{a=7,\cI\cJ} =& \frac{1}{\sqrt2}e^{-\frac{1}{2}(\rho_2+\rho_4)}\frac{1}{\sqrt{\I\,\tau}} \Bigl(\left(T^J_{2I} - \R\,\tau\, T^J_{1I}\right) \eta^{IK} \\& + \tfrac{1}{2}\left(t^1 + \R\,\tau \,t^2\right)\eta^{IJ} \Bigr) E^\cI_K E^\cJ_J \,.
\end{aligned}\end{equation}
For completeness we also give the contracted versions of \(\xi_M\), although they vanish for the special case of the Enriques Calabi-Yau,
\begin{equation}\begin{aligned}
\xi^{m=1} = \xi^{a=6} &= -\frac{1}{\sqrt2}e^{-\frac{1}{2}(\rho_2+\rho_4)}\sqrt{\I\,\tau} \,t^2 \,,\\
\xi^{m=2} = \xi^{a=7} &= \frac{1}{\sqrt2}e^{-\frac{1}{2}(\rho_2+\rho_4)}\frac{1}{\sqrt{\I\,\tau}} \left(t^1 + \R\,\tau \,t^2\right) \,.
\end{aligned}\end{equation}

It is important to notice that these expression are still subject to a set of constraints, since there are redundancies in the scalar sector.
One has to use the relations in \cite{KashaniPoor:2013en} in order to extract the proper unconstrained contracted embedding tensors. For the Enriques Calabi-Yau
we find\footnote{The geometrical analysis of the Enriques Calabi-Yau was also carried out in \cite{KashaniPoor:2013en}.}
\begin{align}
 &f_{1,6\,\,\, 3,8\,\,\, 5,10} = f_{2,7\,\,\, 4,9\,\,\, 5,10} = \frac{1}{2} \Sigma^3 \, e^{-\frac{1}{2}(\rho_2 + \rho_4)}\, \I \, \tau \nn \\
 &\xi_{1,6\,\,\, 3,8} = \xi_{2,7\,\,\, 4,9} = \frac{1}{\sqrt 2}\, e^{-\frac{1}{2}(\rho_2 + \rho_4)}\, \I \, \tau \, ,
\end{align}
where there are two options for each index position.
We explicitly inserted the quantities $t^i_I$, $T^I_{iJ}$ for the Enriques Calabi-Yau \cite{KashaniPoor:2013en}
\begin{align}\label{e:torsion_classes}
 (t^i_I)=&\begin{pmatrix}
 0 & 1 & 0 & 0 & -1 & 0 & \textbf{0}_{1\times 8} \\
 -1 & 0 & 0 & 1 & 0 & 0 & \textbf{0}_{1\times 8}
         \end{pmatrix} \, , \nn\\
 (T^I_{1J})=&\begin{pmatrix}
 0 & 0 & 1 & 0 & 0 & -1 & \textbf{0}_{1\times 8} \\
 0 & 0 & 0 & 0 & 0 & 0 &  \textbf{0}_{1\times 8}\\
 -1 & 0 & 0 & 1 & 0 & 0 & \textbf{0}_{1\times 8} \\
 0 & 0 & 1 & 0 & 0 & -1 & \textbf{0}_{1\times 8} \\
 0 & 0 & 0 & 0 & 0 & 0 & \textbf{0}_{1\times 8} \\
 -1 & 0 & 0 & 1 & 0 & 0 & \textbf{0}_{1\times 8} \\
 \textbf{0}_{8\times 1} & \textbf{0}_{8\times 1} & \textbf{0}_{8\times 1} & \textbf{0}_{8\times 1} & \textbf{0}_{8\times 1} & \textbf{0}_{8\times 1} & 
 \textbf{0}_{8\times 8}
             \end{pmatrix} \, , \nn\\
 (T^I_{2J})=&\begin{pmatrix}
 0 & 0 & 0 & 0 & 0 & 0 & \textbf{0}_{1\times 8} \\
 0 & 0 & 1 & 0 & 0 & -1 & \textbf{0}_{1\times 8}\\
 0 & -1 & 0 & 0 & 1 & 0 & \textbf{0}_{1\times 8} \\
 0 & 0 & 0 & 0 & 0 & 0 & \textbf{0}_{1\times 8} \\
 0 & 0 & 1 & 0 & 0 & -1 & \textbf{0}_{1\times 8} \\
 0 & -1 & 0 & 0 & 1 & 0 & \textbf{0}_{1\times 8} \\
 \textbf{0}_{8\times 1} & \textbf{0}_{8\times 1} & \textbf{0}_{8\times 1} & \textbf{0}_{8\times 1} & \textbf{0}_{8\times 1} & \textbf{0}_{8\times 1} & 
 \textbf{0}_{8\times 8}
             \end{pmatrix} \, .
\end{align}
Note that the general elimination of redundancies is far from being straightforward.

\section{Comparison with type IIA supergravity on \texorpdfstring{$SU(2)$}{SU(2)}-structure manifolds}\label{IIAreduction}
Another way of reproducing the results from \autoref{su2reduction} is to take the four dimensional theory obtained in \cite{KashaniPoor:2013en} by reducing type IIA string theory on \(SU(2)\)-structure manifolds, and relate it to the five dimensional case.
Since type IIA string theory can be obtained from M-theory by compactifying it on a circle, our results should be connected
to the four dimensional theory in the same way.
Thus it is possible to take the dictionary from \cite{Schon:2006kz}, where exactly the relevant compactification of \(\cN = 4\), \(d = 5\) supergravity is described, and uplift the existing results to five dimensions.

It has been worked out in \cite{Danckaert:2011ju} how to group the vectors in four dimensions into \(SO(6,\tilde n)\) representations
\begin{equation}\begin{aligned}
A^{\tilde M+} &= \left(G^i, \tilde{B}^{\bar\imath}, A, \tilde{C}_{12}, C^J \right), \\
A^{\tilde M-} &= \left(B^i, \tilde{G}^{\bar\imath}, C_{12}, \tilde{A}, \tilde{C}^J \right)\,.
\end{aligned}\end{equation}
where \(A^{\tilde M-}\) is the magnetic dual of \(A^{\tilde M+}\)\footnote{
We use indices \(\tilde M, \tilde N, \dots = 1, \dots, 6 + \tilde n\) for the \(SO(6,\tilde n)\) to distinguish them from the \(SO(5,\tilde n - 1)\) indices \(M, N, \dots\).
Notice also that the \(d = 4\) theory contains one additional vector multiplet compared to \(d = 5\), so \(SO(5,\tilde n - 1)\) in five dimensions corresponds indeed to \(SO(6,\tilde n)\) in four dimensions.
}. The \(SO(6,\tilde n)\) metric is given by
\begin{equation}\label{eq:d4metric}
\eta_{\tilde M\tilde N} =
\begin{pmatrix}
0 & \delta_{i\jb} & 0 & 0 & 0 \\
\delta_{\ib j} & 0 & 0 & 0 & 0 \\
0 & 0 & 0 & 1 & 0 \\
0 & 0 & 1 & 0 & 0 \\
0 & 0 & 0 & 0 & \eta_{IJ}
\end{pmatrix}\,.
\end{equation}
It is now necessary to determine how to break \(A^{\tilde M+}\) and \(A^{\tilde M-}\) into \(SO(5,\tilde n-1)\) representations.
Therefore we will write \(\tilde M = \{M, \oplus, \ominus\}\).
Obviously \(A\) does not appear in the five dimensional case.
When tracing back its origin from the reduction of M-theory to IIA supergravity, it is clear that it is the Kaluza-Klein vector, coming from reducing the five dimensional metric to four dimensions. Thus according to \cite{Schon:2006kz} we have to identify it with \(A^{\ominus+}\) and its magnetic dual with\(A^{\oplus-}\).
This makes it furthermore possible to fix \(A^{\oplus+} = \tilde{C}_{12}\) and \(A^{\ominus-} = C_{12}\).
Lastly \(B^i\) and \(\tilde{B}^{\bar\imath}\) do not appear in the five dimensional theory as well, but since they originate from \(C^i\) and \(\tilde{C}^{\bar\imath}\), they can simply be replaced by these.
Using this information, the correct identification of the five dimensional vectors with \(A^M\) and \(A^0\) is
\begin{equation}\begin{aligned}
A^M &= A^{M+} = \left(G^i, \tilde{C}^{\bar\imath}, C^J \right)\,, \\ 
A^0 &= A^{\ominus-} = C_{12}\,,
\end{aligned}\end{equation}
which reproduces the former results. Furthermore we can obtain \eqref{eq:su2metric} by crossing out the fifth and sixth row and line from \eqref{eq:d4metric}.

Note that we can also get \(\Sigma\) and the scalar metric \(M_{MN}\) from the four dimensional results in \cite{Danckaert:2011ju}.
Namely \eqref{eq:scalarmetric} can be obtained from the four dimensional \(M_{\tilde M\tilde N}\) by replacing \(\beta\) with \(\gamma\) and removing all scalars that do not exist in the five dimensional theory. \(\Sigma\) is related to the \(M_{66}\) component
in four dimensions, whereby here the additional factor of \(\I\,\tau\) and the different Weyl rescalings of the metric in four and five dimensions have to be taken into account.

Furthermore in \cite{Schon:2006kz} formulae are provided for the reduction of the embedding tensors,
which together with the expressions from \cite{KashaniPoor:2013en} yield
\begin{equation}\begin{aligned}
\xi_i &= 2 f_{+i\oplus\ominus} = -2 f_{+i56}= - \epsilon_{ij}t^j\,, \\
\xi_{iI} &= f_{-\ominus iI} = f_{-5IJ} = \epsilon_{ij}t^j_I\,, \\
f_{ij\bar\imath} &= f_{+ij\bar\imath} = \delta_{\bar\imath[i}\epsilon_{j]k}t^k\,, \\
f_{iIJ} &= f_{+iIJ} = - T^K_{iI} \eta_{KJ} - \tfrac{1}{2}\epsilon_{ij}t^j \eta_{IJ} \,.
\end{aligned}\end{equation}
For \(\xi_{i}\) one can equally well use the relation
\begin{equation}
\xi_i = \xi_{+i} = - \epsilon_{ij}t^j \,.
\end{equation}

\bibliography{references}

\providecommand{\href}[2]{#2}\begingroup\raggedright\begin{thebibliography}{10}

\bibitem{Cecotti:1984rk}
S.~Cecotti, L.~Girardello, and M.~Porrati, ``{Two into one won't go},'' {\em
  Phys.Lett.} {\bf B145} (1984)
61.

\bibitem{Cecotti:1984wn}
S.~Cecotti, L.~Girardello, and M.~Porrati, ``{Constraints on partial
  superhiggs},'' {\em Nucl.Phys.} {\bf B268} (1986)
295--316.

\bibitem{Cecotti:1985sf}
S.~Cecotti, L.~Girardello, and M.~Porrati, ``{An Exceptional $N=2$ Supergravity
  With Flat Potential and Partial Superhiggs},'' {\em Phys.Lett.} {\bf B168}
  (1986)
83.

\bibitem{Ferrara:1995gu}
S.~Ferrara, L.~Girardello, and M.~Porrati, ``{Minimal Higgs branch for the
  breaking of half of the supersymmetries in N=2 supergravity},'' {\em
  Phys.Lett.} {\bf B366} (1996) 155--159,
\href{http://arXiv.org/abs/hep-th/9510074}{{\tt hep-th/9510074}}.

\bibitem{Antoniadis:1995vb}
I.~Antoniadis, H.~Partouche, and T.~Taylor, ``{Spontaneous breaking of N=2
  global supersymmetry},'' {\em Phys.Lett.} {\bf B372} (1996) 83--87,
\href{http://arXiv.org/abs/hep-th/9512006}{{\tt hep-th/9512006}}.

\bibitem{Fre:1996js}
P.~Fre, L.~Girardello, I.~Pesando, and M.~Trigiante, ``{Spontaneous N=2
  $\rightarrow$ N=1 local supersymmetry breaking with surviving compact gauge
  group},'' {\em Nucl.Phys.} {\bf B493} (1997) 231--248,
\href{http://arXiv.org/abs/hep-th/9607032}{{\tt hep-th/9607032}}.

\bibitem{Kiritsis:1997ca}
E.~Kiritsis and C.~Kounnas, ``{Perturbative and nonperturbative partial
  supersymmetry breaking: N=4 $\rightarrow$ N=2 $\rightarrow$ N=1},'' {\em
  Nucl.Phys.} {\bf B503} (1997) 117--156,
\href{http://arXiv.org/abs/hep-th/9703059}{{\tt hep-th/9703059}}.

\bibitem{Andrianopoli:2002rm}
L.~Andrianopoli, R.~D'Auria, S.~Ferrara, and M.~Lledo, ``{Super Higgs effect in
  extended supergravity},'' {\em Nucl.Phys.} {\bf B640} (2002) 46--62,
\href{http://arXiv.org/abs/hep-th/0202116}{{\tt hep-th/0202116}}.

\bibitem{Louis:2009xd}
J.~Louis, P.~Smyth, and H.~Triendl, ``{Spontaneous N=2 to N=1 Supersymmetry
  Breaking in Supergravity and Type II String Theory},'' {\em JHEP} {\bf 1002}
  (2010) 103,
\href{http://arXiv.org/abs/0911.5077}{{\tt 0911.5077}}.

\bibitem{Louis:2010ui}
J.~Louis, P.~Smyth, and H.~Triendl, ``{The N=1 Low-Energy Effective Action of
  Spontaneously Broken N=2 Supergravities},'' {\em JHEP} {\bf 1010} (2010) 017,
\href{http://arXiv.org/abs/1008.1214}{{\tt 1008.1214}}.

\bibitem{Hansen:2013dda}
T.~Hansen and J.~Louis, ``{Examples of $\mathcal{N} =$ 2 to $\mathcal{N} =$ 1
  supersymmetry breaking},'' {\em JHEP} {\bf 1311} (2013) 075,
\href{http://arXiv.org/abs/1306.5994}{{\tt 1306.5994}}.

\bibitem{Hohm:2004rc}
O.~Hohm and J.~Louis, ``{Spontaneous N = 2 $\rightarrow$ N = 1 supergravity
  breaking in three- dimensions},'' {\em Class.Quant.Grav.} {\bf 21} (2004)
  4607--4624,
\href{http://arXiv.org/abs/hep-th/0403128}{{\tt hep-th/0403128}}.

\bibitem{Grimm:2014soa}
T.~W. Grimm and A.~Kapfer, ``{Self-Dual Tensors and Partial Supersymmetry
  Breaking in Five Dimensions},''
\href{http://arXiv.org/abs/1402.3529}{{\tt 1402.3529}}.

\bibitem{Schon:2006kz}
J.~Schon and M.~Weidner, ``{Gauged N=4 supergravities},'' {\em JHEP} {\bf 0605}
  (2006) 034,
\href{http://arXiv.org/abs/hep-th/0602024}{{\tt hep-th/0602024}}.

\bibitem{Danckaert:2011ju}
T.~Danckaert, J.~Louis, D.~Martinez-Pedrera, B.~Spanjaard, and H.~Triendl,
  ``{The N=4 effective action of type IIA supergravity compactified on
  SU(2)-structure manifolds},'' {\em JHEP} {\bf 1108} (2011) 024,
\href{http://arXiv.org/abs/1104.5174}{{\tt 1104.5174}}.

\bibitem{KashaniPoor:2013en}
A.-K. Kashani-Poor, R.~Minasian, and H.~Triendl, ``{Enhanced supersymmetry from
  vanishing Euler number},'' {\em JHEP} {\bf 1304} (2013) 058,
\href{http://arXiv.org/abs/1301.5031}{{\tt 1301.5031}}.

\bibitem{Cadavid:1995bk}
A.~Cadavid, A.~Ceresole, R.~D'Auria, and S.~Ferrara, ``{Eleven-dimensional
  supergravity compactified on Calabi-Yau threefolds},'' {\em Phys.Lett.} {\bf
  B357} (1995) 76--80,
\href{http://arXiv.org/abs/hep-th/9506144}{{\tt hep-th/9506144}}.

\bibitem{Papadopoulos:1995da}
G.~Papadopoulos and P.~Townsend, ``{Compactification of D = 11 supergravity on
  spaces of exceptional holonomy},'' {\em Phys.Lett.} {\bf B357} (1995)
  300--306,
\href{http://arXiv.org/abs/hep-th/9506150}{{\tt hep-th/9506150}}.

\bibitem{Ferrara:1996hh}
S.~Ferrara, R.~R. Khuri, and R.~Minasian, ``{M theory on a Calabi-Yau
  manifold},'' {\em Phys.Lett.} {\bf B375} (1996) 81--88,
\href{http://arXiv.org/abs/hep-th/9602102}{{\tt hep-th/9602102}}.

\bibitem{Cassani:2010uw}
D.~Cassani, G.~Dall'Agata, and A.~F. Faedo, ``{Type IIB supergravity on
  squashed Sasaki-Einstein manifolds},'' {\em JHEP} {\bf 1005} (2010) 094,
\href{http://arXiv.org/abs/1003.4283}{{\tt 1003.4283}}.

\bibitem{Liu:2010sa}
J.~T. Liu, P.~Szepietowski, and Z.~Zhao, ``{Consistent massive truncations of
  IIB supergravity on Sasaki-Einstein manifolds},'' {\em Phys.Rev.} {\bf D81}
  (2010) 124028,
\href{http://arXiv.org/abs/1003.5374}{{\tt 1003.5374}}.

\bibitem{Gauntlett:2010vu}
J.~P. Gauntlett and O.~Varela, ``{Universal Kaluza-Klein reductions of type IIB
  to N=4 supergravity in five dimensions},'' {\em JHEP} {\bf 1006} (2010) 081,
\href{http://arXiv.org/abs/1003.5642}{{\tt 1003.5642}}.

\bibitem{Tsikas:1986rx}
T.~Tsikas, ``{Consistent Truncations of Chiral $N=2 D=10$ Supergravity on the
  Round Five Sphere},'' {\em Class.Quant.Grav.} {\bf 3} (1986)
733.

\bibitem{Lu:1999bw}
H.~Lu, C.~Pope, and T.~A. Tran, ``{Five-dimensional N=4, SU(2) x U(1) gauged
  supergravity from type IIB},'' {\em Phys.Lett.} {\bf B475} (2000) 261--268,
  \href{http://arXiv.org/abs/hep-th/9909203}{{\tt hep-th/9909203}}.

\bibitem{Ceresole:1999zs}
A.~Ceresole, G.~Dall'Agata, R.~D'Auria, and S.~Ferrara, ``{Spectrum of type IIB
  supergravity on AdS(5) x T**11: Predictions on N=1 SCFT's},'' {\em Phys.Rev.}
  {\bf D61} (2000) 066001,
\href{http://arXiv.org/abs/hep-th/9905226}{{\tt hep-th/9905226}}.

\bibitem{Cvetic:2000nc}
M.~Cvetic, H.~Lu, C.~Pope, A.~Sadrzadeh, and T.~A. Tran, ``{Consistent SO(6)
  reduction of type IIB supergravity on S**5},'' {\em Nucl.Phys.} {\bf B586}
  (2000) 275--286,
\href{http://arXiv.org/abs/hep-th/0003103}{{\tt hep-th/0003103}}.

\bibitem{Hoxha:2000jf}
P.~Hoxha, R.~Martinez-Acosta, and C.~Pope, ``{Kaluza-Klein consistency, Killing
  vectors, and Kahler spaces},'' {\em Class.Quant.Grav.} {\bf 17} (2000)
  4207--4240,
\href{http://arXiv.org/abs/hep-th/0005172}{{\tt hep-th/0005172}}.

\bibitem{Buchel:2006gb}
A.~Buchel and J.~T. Liu, ``{Gauged supergravity from type IIB string theory on
  Y**p,q manifolds},'' {\em Nucl.Phys.} {\bf B771} (2007) 93--112,
\href{http://arXiv.org/abs/hep-th/0608002}{{\tt hep-th/0608002}}.

\bibitem{Gauntlett:2007ma}
J.~P. Gauntlett and O.~Varela, ``{Consistent Kaluza-Klein reductions for
  general supersymmetric AdS solutions},'' {\em Phys.Rev.} {\bf D76} (2007)
  126007,
\href{http://arXiv.org/abs/0707.2315}{{\tt 0707.2315}}.

\bibitem{Skenderis:2010vz}
K.~Skenderis, M.~Taylor, and D.~Tsimpis, ``{A Consistent truncation of IIB
  supergravity on manifolds admitting a Sasaki-Einstein structure},'' {\em
  JHEP} {\bf 1006} (2010) 025,
\href{http://arXiv.org/abs/1003.5657}{{\tt 1003.5657}}.

\bibitem{Cassani:2010na}
D.~Cassani and A.~F. Faedo, ``{A Supersymmetric consistent truncation for
  conifold solutions},'' {\em Nucl.Phys.} {\bf B843} (2011) 455--484,
\href{http://arXiv.org/abs/1008.0883}{{\tt 1008.0883}}.

\bibitem{Bena:2010pr}
I.~Bena, G.~Giecold, M.~Grana, N.~Halmagyi, and F.~Orsi, ``{Supersymmetric
  Consistent Truncations of IIB on $T^{1,1}$},'' {\em JHEP} {\bf 1104} (2011)
  021,
\href{http://arXiv.org/abs/1008.0983}{{\tt 1008.0983}}.

\bibitem{Bah:2010cu}
I.~Bah, A.~Faraggi, J.~I. Jottar, and R.~G. Leigh, ``{Fermions and Type IIB
  Supergravity On Squashed Sasaki-Einstein Manifolds},'' {\em JHEP} {\bf 1101}
  (2011) 100,
\href{http://arXiv.org/abs/1009.1615}{{\tt 1009.1615}}.

\bibitem{Liu:2010pq}
J.~T. Liu, P.~Szepietowski, and Z.~Zhao, ``{Supersymmetric massive truncations
  of IIb supergravity on Sasaki-Einstein manifolds},'' {\em Phys.Rev.} {\bf
  D82} (2010) 124022,
\href{http://arXiv.org/abs/1009.4210}{{\tt 1009.4210}}.

\bibitem{Liu:2011dw}
J.~T. Liu and P.~Szepietowski, ``{Supersymmetry of consistent massive
  truncations of IIB supergravity},'' {\em Phys.Rev.} {\bf D85} (2012) 126010,
\href{http://arXiv.org/abs/1103.0029}{{\tt 1103.0029}}.

\bibitem{Halmagyi:2011yd}
N.~Halmagyi, J.~T. Liu, and P.~Szepietowski, ``{On N = 2 Truncations of IIB on
  $T^{1,1}$},'' {\em JHEP} {\bf 1207} (2012) 098,
\href{http://arXiv.org/abs/1111.6567}{{\tt 1111.6567}}.

\bibitem{Gauntlett:2004zh}
J.~P. Gauntlett, D.~Martelli, J.~Sparks, and D.~Waldram, ``{Supersymmetric
  AdS(5) solutions of M theory},'' {\em Class.Quant.Grav.} {\bf 21} (2004)
  4335--4366,
\href{http://arXiv.org/abs/hep-th/0402153}{{\tt hep-th/0402153}}.

\bibitem{Gauntlett:2006ai}
J.~P. Gauntlett, E.~O~Colgain, and O.~Varela, ``{Properties of some conformal
  field theories with M-theory duals},'' {\em JHEP} {\bf 0702} (2007) 049,
\href{http://arXiv.org/abs/hep-th/0611219}{{\tt hep-th/0611219}}.

\bibitem{Gauntlett:2007sm}
J.~P. Gauntlett and O.~Varela, ``{D=5 SU(2) x U(1) Gauged Supergravity from
  D=11 Supergravity},'' {\em JHEP} {\bf 0802} (2008) 083,
\href{http://arXiv.org/abs/0712.3560}{{\tt 0712.3560}}.

\bibitem{OColgain:2011ng}
E.~O~Colgain and O.~Varela, ``{Consistent reductions from D=11 beyond
  Sasaki-Einstein},'' {\em Phys.Lett.} {\bf B703} (2011) 180--185,
\href{http://arXiv.org/abs/1106.4781}{{\tt 1106.4781}}.

\bibitem{Sfetsos:2014tza}
K.~Sfetsos and D.~C. Thompson, ``{New ${\cal N} = 1$ supersymmetric $AdS_5$
  backgrounds in Type IIA supergravity},''
\href{http://arXiv.org/abs/1408.6545}{{\tt 1408.6545}}.

\bibitem{Dall'Agata:2001vb}
G.~Dall'Agata, C.~Herrmann, and M.~Zagermann, ``{General matter coupled N=4
  gauged supergravity in five-dimensions},'' {\em Nucl.Phys.} {\bf B612} (2001)
  123--150,
\href{http://arXiv.org/abs/hep-th/0103106}{{\tt hep-th/0103106}}.

\bibitem{Cassani:2012wc}
D.~Cassani, G.~Dall'Agata, and A.~F. Faedo, ``{BPS domain walls in N=4
  supergravity and dual flows},'' {\em JHEP} {\bf 1303} (2013) 007,
\href{http://arXiv.org/abs/1210.8125}{{\tt 1210.8125}}.

\bibitem{Hitchin:2000sk}
N.~J. Hitchin, ``{The Geometry of three forms in six-dimensions},'' {\em
  J.Diff.Geom.} {\bf 55} (2000) 547--576,
\href{http://arXiv.org/abs/math.DG/0010054}{{\tt math.DG/0010054}}.

\bibitem{hitchin2001stable}
N.~Hitchin, ``Stable forms and special metrics,'' {\em Contemporary
  Mathematics} {\bf 288} (2001) 70--89,
  \href{http://arXiv.org/abs/math.DG/0107101}{{\tt math.DG/0107101}}.

\bibitem{chiossi2002intrinsic}
S.~Chiossi and S.~Salamon, ``The intrinsic torsion of SU (3) and G2
  structures,'' {\em Differential Geometry, Valencia 2001} (2002) 115--133,
  \href{http://arXiv.org/abs/math.DG/0202282}{{\tt math.DG/0202282}}.

\bibitem{Gauntlett:2002sc}
J.~P. Gauntlett, D.~Martelli, S.~Pakis, and D.~Waldram, ``{G structures and
  wrapped NS5-branes},'' {\em Commun.Math.Phys.} {\bf 247} (2004) 421--445,
\href{http://arXiv.org/abs/hep-th/0205050}{{\tt hep-th/0205050}}.

\bibitem{Gauntlett:2003cy}
J.~P. Gauntlett, D.~Martelli, and D.~Waldram, ``{Superstrings with intrinsic
  torsion},'' {\em Phys.Rev.} {\bf D69} (2004) 086002,
\href{http://arXiv.org/abs/hep-th/0302158}{{\tt hep-th/0302158}}.

\bibitem{Bovy:2005qq}
J.~Bovy, D.~Lust, and D.~Tsimpis, ``{N = 1,2 supersymmetric vacua of IIA
  supergravity and SU(2) structures},'' {\em JHEP} {\bf 0508} (2005) 056,
\href{http://arXiv.org/abs/hep-th/0506160}{{\tt hep-th/0506160}}.

\bibitem{ReidEdwards:2008rd}
R.~Reid-Edwards and B.~Spanjaard, ``{N=4 Gauged Supergravity from Duality-Twist
  Compactifications of String Theory},'' {\em JHEP} {\bf 0812} (2008) 052,
\href{http://arXiv.org/abs/0810.4699}{{\tt 0810.4699}}.

\bibitem{Lust:2009zb}
D.~Lust and D.~Tsimpis, ``{Classes of AdS(4) type IIA/IIB compactifications
  with SU(3) x SU(3) structure},'' {\em JHEP} {\bf 0904} (2009) 111,
\href{http://arXiv.org/abs/0901.4474}{{\tt 0901.4474}}.

\bibitem{Triendl:2009ap}
H.~Triendl and J.~Louis, ``{Type II compactifications on manifolds with SU(2) x
  SU(2) structure},'' {\em JHEP} {\bf 0907} (2009) 080,
\href{http://arXiv.org/abs/0904.2993}{{\tt 0904.2993}}.

\bibitem{Louis:2009dq}
J.~Louis, D.~Martinez-Pedrera, and A.~Micu, ``{Heterotic compactifications on
  SU(2)-structure backgrounds},'' {\em JHEP} {\bf 0909} (2009) 012,
\href{http://arXiv.org/abs/0907.3799}{{\tt 0907.3799}}.

\bibitem{Bonetti:2013ela}
F.~Bonetti, T.~W. Grimm, and S.~Hohenegger, ``{One-loop Chern-Simons terms in
  five dimensions},'' {\em JHEP} {\bf 1307} (2013) 043,
\href{http://arXiv.org/abs/1302.2918}{{\tt 1302.2918}}.

\bibitem{Bonetti:2012fn}
F.~Bonetti, T.~W. Grimm, and S.~Hohenegger, ``{A Kaluza-Klein inspired action
  for chiral p-forms and their anomalies},'' {\em Phys.Lett.} {\bf B720} (2013)
  424--427,
\href{http://arXiv.org/abs/1206.1600}{{\tt 1206.1600}}.

\bibitem{Bonetti:2011mw}
F.~Bonetti and T.~W. Grimm, ``{Six-dimensional (1,0) effective action of
  F-theory via M-theory on Calabi-Yau threefolds},'' {\em JHEP} {\bf 1205}
  (2012) 019,
\href{http://arXiv.org/abs/1112.1082}{{\tt 1112.1082}}.

\bibitem{Grimm:2013oga}
T.~W. Grimm, A.~Kapfer, and J.~Keitel, ``{Effective action of 6D F-Theory with
  U(1) factors: Rational sections make Chern-Simons terms jump},'' {\em JHEP}
  {\bf 1307} (2013) 115,
\href{http://arXiv.org/abs/1305.1929}{{\tt 1305.1929}}.

\bibitem{Intriligator:1997pq}
K.~A. Intriligator, D.~R. Morrison, and N.~Seiberg, ``{Five-dimensional
  supersymmetric gauge theories and degenerations of Calabi-Yau spaces},'' {\em
  Nucl.Phys.} {\bf B497} (1997) 56--100,
\href{http://arXiv.org/abs/hep-th/9702198}{{\tt hep-th/9702198}}.

\end{thebibliography}\endgroup
\bibliographystyle{utcaps} 
\end{document}